\documentclass{emulateapj}
\usepackage{epstopdf}
\usepackage{CJK}
\usepackage{color}
\usepackage{pshan}
\newcommand{\beq}{\begin{equation}}
\newcommand{\eeq}{\end{equation}}
\newcommand{\bea}{\begin{eqnarray}}
\newcommand{\eea}{\end{eqnarray}}

\begin{document}
%\title{The Nonlinear, Cosmological Transverse-Momentum Power Spectrum of the Intergalactic Medium and the Kinetic Sunyaev-Zel'dovich Effect}
\begin{CJK}[HL]{KS}{}
\title{The Impact of Nonlinear Structure Formation on the Power Spectrum of Transverse Momentum Fluctuations and the Kinetic Sunyaev-Zel'dovich Effect}
%\title{The Impact of Nonlinear Structure Formation on the Transverse Momentum Power Spectrum and the Kinetic Sunyaev-Zel'dovich Effect}

\author{Hyunbae Park (¹ÚÇö¹è)\altaffilmark{1,2}}
\author{Eiichiro Komatsu\altaffilmark{3,4}}
\author{Paul R. Shapiro\altaffilmark{1}}
\author{Jun Koda\altaffilmark{5}}
\author{Yi Mao\altaffilmark{6,7}}

\altaffiltext{1}{Texas Cosmology Center and the Department of Astronomy, The University of Texas at Austin, 1 University Station, C1400, Austin, TX 78712, USA}
\altaffiltext{2}{Korea Astronomy and Space Science Institute, Daejeon, 305-348, Korea}
\altaffiltext{3}{Kavli Institute for the Physics and Mathematics of the Universe, Todai Institutes for Advanced Study, the University of Tokyo, Kashiwa, Japan 277-8583 (Kavli IPMU, WPI)}
\altaffiltext{4}{Max-Planck-Institut f\"{u}r Astrophysik, Karl-Schwarzschild Str. 1, 85741 Garching, Germany}
\altaffiltext{5}{Osservatorio Astronomico di Brera, via E. Bianchi 46, 23807 Merate, Italy}
\altaffiltext{6}{Department of Physics and Tsinghua Center for Astrophysics, Tsinghua University, Beijing 100084, China}
\altaffiltext{7}{Sorbonne Universit\'es, UPMC Univ Paris 6 and CNRS, UMR 7095, Institut d'Astrophysique de Paris, Institut Lagrange de Paris, 98 bis bd Arago, 75014 Paris, France}

\begin{abstract}
Cosmological {\it transverse} momentum fields, whose directions are
 perpendicular to Fourier wave vectors, induce temperature anisotropies
 in the cosmic microwave background via the kinetic Sunyaev-Zel'dovich
 (kSZ) effect.
 The transverse momentum power spectrum contains the
 four-point function of density and velocity fields,
 $\langle\delta\delta v v\rangle$. In the post-reionization epoch,
 nonlinear effects dominate in the power spectrum. We use
 perturbation theory and cosmological $N$-body simulations to
 calculate this nonlinearity. We derive the next-to-leading order
 expression for the power spectrum with a particular emphasis on
 the connected term that has been ignored in the literature. While the
 contribution from the connected term on small scales
 ($k>0.1~h~\rm{Mpc}^{-1}$) is subdominant relative to the unconnected
 term, we find that its contribution to the kSZ
 power spectrum at $\ell = 3000$ at $z<6$ can be as large as ten percent
 of the unconnected term, which would reduce the allowed contribution
 from the reionization epoch ($z>6$) by twenty percent. The power spectrum of transverse momentum on large scales is expected to scale as $k^2$ as a consequence of momentum conservation. We show that both the leading and the next-to-leading order terms satisfy this scaling. In particular, we find that both of the unconnected and connected terms are necessary to reproduce $k^2$.
\end{abstract}

\section{Introduction}

The cosmic transverse momentum field is observable in temperature
fluctuations of the cosmic microwave background (CMB). The line-of-sight
(LOS) component of the momentum of
electrons in the ionized intergalactic medium (IGM) induces temperature fluctuations via the Doppler effect,
and this is known as the kinetic Sunyaev-Zel'dovich (kSZ) effect
\citep{zeldovich/sunyaev:1969}. The contribution from the longitudinal
momentum field, whose direction is parallel to the wave vector, suffers
from cancellation of the contributions of troughs and crests integrated
along the LOS; thus, the contribution from the transverse momentum field
dominates \citep{vishniac:1987}.

The kSZ effect is given by  \citep{sunyaev/zeldovich:1980}
\beq \label{Eq:kSZ}
\frac{\Delta T (\hat\gamma)}{T}= -\int d\tau e^{-\tau} \frac{\hat{\gamma}\cdot\bold{v}}{c},
\eeq
where $\hat{\gamma}$ is the LOS unit vector, $\bold{v}$ the
peculiar velocity field, and $\tau$ the optical depth to Thomson
scattering integrated through the IGM from $z=0$ to the scatterer. The
differential form of $\tau$ is
$d\tau = c~n_e \sigma_{\rm{T}}\left(\frac{dt}{dz}\right)dz$, where $n_e$ is
the number density of free electrons in the IGM, and $\sigma_{\rm{T}}$
is the Thomson scattering cross section. 

Equation~(\ref{Eq:kSZ}) can be rewritten in the following form:
\beq \label{Eq:kSZ2}
\frac{\Delta T}{T}(\hat{\gamma}) = -\frac{\sigma_T \bar{n}_{e,0}}{c}\int \frac{ds}{a^2} ~ e^{-\tau} \bold{q}\cdot \hat{\gamma},
\eeq
where $\bold{q} \equiv \chi \bold{v}(1+\delta)$ is the momentum of
ionized gas, $\delta$ is the fractional mass density perturbation of the gas,
$\chi \equiv n_e/(n_{\rm{H}}+2n_{\rm{He}})$ the
ionization fraction, $\bar{n}_{e,0}
= \bar{n}_{\rm{H,0}}+2\bar{n}_{\rm{He,0}}$ the mean number density
of electrons at the (fully-ionized) present epoch, and $s$ the
distance photons travelled from a source to the observer in comoving
units.  

Longitudinal momentum fields cancel out in the line-of-sight integral of Equation~(\ref{Eq:kSZ2}) \cite[See Appendix~\ref{sec:P_q,par} for a quantitative argument; also see][]{vishniac:1987}.
%As longitudinal momentum fields cancel out in the line-of-sight integral of Equation~(\ref{Eq:kSZ2}),
Then, the angular power spectrum of Equation~(\ref{Eq:kSZ2}), $C_\ell$, at
large multipoles is dominated by the power spectrum of the transverse momentum field, $P_{q_{\perp}}(k)$, and is given by \citep{vishniac:1987}
\beq \label{C_l}
C_\ell = \frac12
\left(\frac{\sigma_T \bar{n}_{e,0}}{c}  \right)^2
\int\frac{ds}{s^2a^4} e^{-2\tau}
P_{q_{\perp}}\left(k=\frac{l}{s},s\right).
\eeq
The kSZ power spectrum has been constrained by observations.
The South Pole Telescope (SPT) and Atacama Cosmology Telescope (ACT)
measure CMB temperature anisotropy at scales beyond the Silk damping
scale ($\ell\gtrsim3000$) where it is possible to distinguish the kSZ
signal from the primary signal. The latest constraint is
$D^{\rm{kSZ}}_{\ell=3000}=2.9\pm 1.3~\mu K^2$ (69\%~CL), where
$D_\ell\equiv \ell(\ell+1)C_\ell/2\pi$ \citep{2014arXiv1408.3161G}.

The measured kSZ power spectrum is the sum of contributions from the
epoch of reionization (EoR) and the post-reionization
epoch. The model of \cite{shaw/etal:2011} suggests that the latter gives
$D^{\rm{kSZ}}_{\ell=3000} \approx 2.0~\mu K^2$, assuming that the
universe became fully ionized at $z=6$. Thus, the central value of the
current measurement implies that the post-reionization kSZ signal
is a factor of two greater than the EoR signal. This highlights the
importance of understanding the post-reionization signal; if we
mis-calculate the post-reionization signal by ten percent, the EoR
signal would be mis-estimated by twenty percent. This motivates our
revisiting assumptions used by the previous calculations of the
post-reionization kSZ power spectrum.

Inhomogeneity in the ionization fraction during the EoR gives a
large boost in the kSZ power spectrum compared to homogeneous ionization
\citep{2013ApJ...769...93P}. The physics that determines this
inhomogeneity is complex, and many papers have been written on this
subject, ranging from early analytical calculations
\citep{gruzinov/hu:1998,santos/etal:2003} and semi-numerical calculations
using an analytical ansatz applied to $N$-body simulations
\citep{zahn/etal:2005,mcquinn/etal:2005,mesinger/mcquinn/spergel:2011,zahn/etal:2011,battaglia/etal:2012b},
to fully numerical simulations that couple the cosmological structure
formation with radiative transfer
\citep{iliev/etal:2007,2013ApJ...769...93P}. 

Modeling the post-reionization signal is simpler because the IGM is
fully ionized and $\chi$ does not fluctuate to a good approximation. We
thus need to model the density and velocity fluctuations of gas. When
$\chi=1$, the momentum field is given by $\bold{q}
=\bold{v}(1+\delta)$. Since the velocity field is purely longitudinal in
the linear regime, the transverse momentum field, $\bold{q}_\perp$, is given by
$\bold{q}_\perp=(\bold{v}\delta)_\perp$ at leading order. The power spectrum is then
given by the four-point function of two densities and
velocities. Schematically, it is given by $\langle qq\rangle = \langle
vv\rangle\langle \delta\delta\rangle+2\langle v\delta\rangle^2+\langle
v\delta v\delta\rangle_{c}$. The last term is called the connected
four-point term, while the others are the unconnected ones. 

In the previous work, the connected term has been ignored. For example, the
earlier analytical studies ignore nonlinearity in density or velocity, and
calculate only the unconnected terms using linear perturbation theory
\citep{ostriker/vishniac:1986,vishniac:1987,dodelson/jubas:1995,jaffe/kamionkowski:1998}. An
analytical model of \cite{hu:2000} is still based upon
 linear theory for velocity, and ignores the connected term,
but replaces the linear density power spectrum with a model for the nonlinear
density power spectrum by \cite{peacock/dodds:1996}.
\cite{ma/fry:2002} use a similar approach with a halo model for the
nonlinear density power spectrum and argue that the connected term is
negligible at large $k$. While we broadly agree with this
conclusion, our aim is to quantify the contribution of the connected
term at large $k$, and also clarify the role of the connected
term in obtaining the correct small-$k$ limit of the transverse
momentum power spectrum in perturbation theory.

Some of the previous ``numerical'' calculations of the post-reionization kSZ
power spectrum \citep{2004MNRAS.347.1224Z,shaw/etal:2011} still rely on
the above analytical model that ignores the connected term, but
takes the ingredient of the model, i.e., nonlinear gas density
power spectrum, from hydrodynamical simulations. Therefore,
quantifying better the contribution of the connected term affects the
results from the previous numerical work as
well. \cite{springel/white/hernquist:2001} and \cite{2001MNRAS.326..155D} computed the kSZ power
spectrum directly from their simulation and did not rely on the model.

Throughout this paper, we shall assume that gas traces dark matter. This
is not a great approximation: shocks in the IGM generated by structure
formation heat gas to high temperatures
\citep[e.g.,][]{cen/ostriker:1999}, and the resulting gas pressure makes
gas less clustered than dark matter particles. 
This effect on the kSZ power spectrum is modest at $\ell=3000$ \citep{shaw/etal:2011,hu:2000}. Star formation converts gas into stars, further reducing the kSZ effect. Shaw et al. (2012) find that $D^{\rm kSZ}_{\ell=3000}$ can be suppressed by as much as 33\% of the simulation without gas cooling and star formation.
Our goal in this paper is to
quantify the error we make by ignoring the connected four-point term in
the transverse momentum power spectrum. While our dark-matter-only
results cannot be extrapolated to gas, we expect that a similar
conclusion would still apply to gas, at least qualitatively.

The remainder of this paper is organized as follows. In
Section~\ref{sec:simulation}, we briefly introduce the $N$-body
simulation we use in this paper. In Section~\ref{sec:basics}, we review
the derivation of the current nonlinear transverse momentum power
spectrum model \citep{hu:2000,ma/fry:2002}.
In Section~\ref{sec:results}, we first confirm that the current model
accurately approximates the unconnected term. We then show that
our simulation data of the transverse momentum power spectrum exceed the model. We argue that this excess is due
to the connected term, by showing that the perturbation theory calculation of
the next-to-leading order terms (including the connected term) explains
the excess at quasi-linear scales successfully. We also show that the
connected term is essential in obtaining the correct low-$k$ limit of
the transverse momentum power spectrum in perturbation theory.
In Section~\ref{sec:PRKSZ}, we quantify the impact of the connected term on the
post-reionization kSZ power spectrum. We conclude in Section~\ref{sec:Conclusion}.

\section{Simulation} \label{sec:simulation}
We shall use a cosmological $N$-body simulation of collisionless particles using 
the ``CubeP$^3$M'' $N$-body code \citep{2013MNRAS.436..540H}. The
simulation is run with $3456^3$ particles in a comoving box of
$1~h^{-1}~{\rm Gpc}$ on a side and is started at $z=150$ using the initial condition generated 
using the Zel'dovich approximation and initial density power spectrum from
 the publicly available code CAMB \citep{2000ApJ...538..473L}.
This simulation was previously presented in \citet{2013MNRAS.433.1230W}.

The resolution of this simulation allows us to sample on average
$\sim 40$ particles per $({\rm Mpc}/h)^3$. This resolution allows us to
avoid sampling artifacts in the velocity power spectrum up to
$k\sim1~h~\rm{Mpc}^{-1}$ \citep[see Figure 3 of][]{2015PhRvD..91d3522Z}.
We then adaptively smooth particles to a grid of $(1008)^3$ cells
\footnote{This adaptive smoothing is by an approach similar to that used in
Smoothed Particle Hydrodynamics.  In this case, spherical smoothing
kernels are assigned to each particle, with their smoothing lengths
adjusted so as to enclose the locations of the 32 nearest-neighbor
particles.  The mass per particle assigned to a given grid cell
then corresponds to the fraction of its finite kernel volume which
overlaps the cell volume.}. 
Therefore, our simulation covers a dynamic range of $0.006\lesssim k\lesssim1~h~\rm{Mpc}^{-1}$ in wavenumber.  The background cosmology is based on the {\sl WMAP} 5-year data combined with constraints from baryonic acoustic oscillations, from observations of galaxies and large-scale structures, and from high-redshift Type Ia supernovae \citep[$\Omega_M = 0.279, \Omega_\Lambda  = 0.721, h =
0.701, \Omega_b = 0.0462, \sigma_8 = 0.8, n_s = 0.96$;][]{komatsu/etal:2009}. 

\section{Transverse momentum power spectrum} \label{sec:basics}

\subsection{Basics}
In the post-reionization era, helium atoms are singly ionized until He
II reionization occurs. We assume that hydrogen reionization finished at
$z=6$ and He II reionization occurred instantaneously at $z=3$; thus,
$\chi=0.93$ for $3<z<6$ and $\chi=1$ for $z\leq 3$. For the rest of the
paper, we shall drop $\chi$ for notational simplicity and write
$\bold{q}=\bold{v}(1+\delta)$. Therefore, the momentum power spectrum
below should be rescaled by $\chi^2$ to yield the correct value needed
for computing the kSZ power spectrum in Section~\ref{sec:PRKSZ}.

\begin{figure}
  \begin{center}
    \includegraphics[scale=0.63]{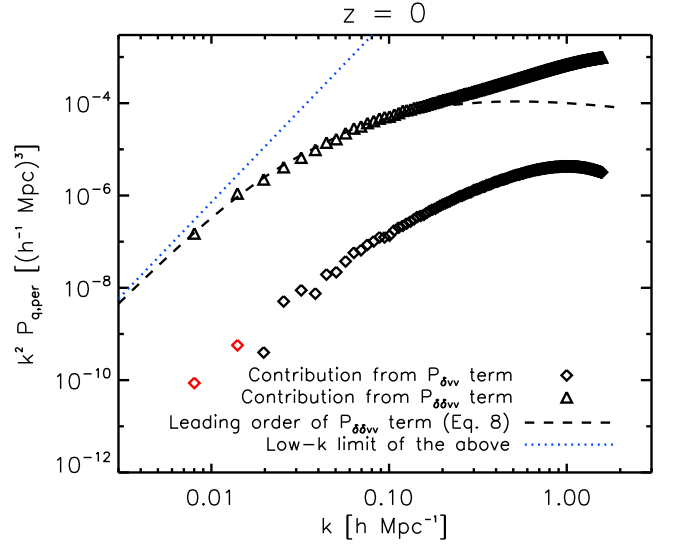}
  \caption{Contributions to $k^2P_{q_\perp}$ from $\left< \delta vv
   \right>$ (diamonds) and $\langle\delta\delta vv\rangle$ (triangles) in Equation~(\ref{eq:Pq_brac})
   measured from our simulation at $z=0$. The red triangles show
   negative values.
   The dashed line shows the
   lowest order calculation for the $\left< \delta\delta vv\right>$
   term, while the dotted line shows its low-$k$ limit given in Equation~(\ref{eq:P_OV^{lowk}}).}
  \label{fig:P_q_sim}
  \end{center}
\end{figure}

We start by Fourier transforming the momentum field,
$\bold{q}=\bold{v}(1+\delta)$, as
\bea 
\bold{\tilde q(k)} = \bold{\tilde v(k)} + \int \frac{d^3 k^\prime}{(2\pi)^3} \tilde\delta\bold{(k-k^\prime)}\bold{\tilde v(k^\prime)},
\eea
where $\tilde\bold{q}(\bold{k})\equiv \int d^3\bold{x}~e^{i\bold{k}\cdot
\bold{x}} \bold{q}(\bold{x})$, etc. Then,
its power spectrum is defined by
\bea 
&&(2\pi)^3 P^{ij}_{qq}(\bold{k_1}) \delta_D(\bold{k_1}+\bold{k_2})
\equiv \left<\tilde q^i (\bold{k_1}) \tilde q^j (\bold{k_2}) \right>
\nonumber\\
&=&  
\left<\tilde v^i (\bold{k_1}) \tilde v^j (\bold{k_2}) \right>+
2 \int \frac{d^3 k^\prime}{(2\pi)^3} \left<
\tilde\delta (\bold{k_1}-\bold{k^\prime})
\tilde v^i (\bold{k^\prime})
\tilde v^j (\bold{k_2}) 
\right>
\nonumber\\ 
&+& 
\int \frac{d^3 k^\prime}{(2\pi)^3}
\int \frac{d^3 k^{\prime\prime}}{(2\pi)^3}
\nonumber
\left<
\tilde\delta (\bold{k_1}-\bold{k^\prime})
\tilde v^i (\bold{k^\prime})
\tilde\delta (\bold{k_2}-\bold{k^{\prime\prime}})
\tilde v^j (\bold{k^{\prime\prime}}) 
\right>,\\
\label{eq:Pq_brac}
\eea 
where the indices, $i$ and $j$, denote the $i$'th and $j$'th components of the vector, respectively. The power spectrum of the transverse mode, $\tilde\bold{q}_\perp \equiv  \tilde\bold{q} -\hat\bold{k}[\tilde\bold{q} \cdot \hat\bold{k}]$, is given by $P_{q_\perp}\equiv\Sigma_i P^{ii}_{qq} [1-(\hat k^{i})^2 ]$. 
In linear and quasi-linear regimes ($k\lesssim 1~h/\rm{Mpc}$), the velocity field is longitudinal (i.e. $\tilde\bold{v}(\bold{k})=\hat\bold{k} \tilde{v}$) to a good approximation, which implies that 
$\left< vv\right>$ and $\left< \delta vv\right>$ in Equation~(\ref{eq:Pq_brac})
vanish. Figure~\ref{fig:P_q_sim} shows that the contribution to
$P_{q_\perp}$ from  $\langle \delta vv
\rangle$ (diamonds) is a few orders of magnitude smaller than
$\langle\delta \delta v v\rangle$ (triangles). The contribution from
$\langle vv\rangle$ (not shown) is even smaller than $\langle \delta vv
\rangle$ by another two orders of magnitude.

Keeping only the last term in Equation~(\ref{eq:Pq_brac}), we obtain
\bea \label{eq:Pq}
P_{q_\perp}(k) &=& 
\int \frac{d^3 k^\prime}{(2\pi)^3}
\int \frac{d^3 k^{\prime\prime}}{(2\pi)^3}
[  \hat\bold{k^{\prime}}\cdot  \hat\bold{k^{\prime\prime}} - (\hat\bold{k} \cdot \hat\bold{k^{\prime}}) (\hat\bold{k} \cdot \hat\bold{k^{\prime\prime}}) ]
\nonumber\\ 
&&
P_{\delta \delta v v} (\bold{k}-\bold{k^\prime},-\bold{k}-\bold{k^{\prime\prime}},\bold{k^\prime},\bold{k^{\prime\prime}}),
\eea 
where $P_{\delta \delta v v}$ is defined by
\bea
&&(2\pi)^3P_{\delta \delta v v}(\bold{k_1},\bold{k_2},\bold{k_3},\bold{k_4})
\delta_D(\bold{k_1}+\bold{k_2}+\bold{k_3}+\bold{k_4}) 
\nonumber \\
&\equiv& \left<
\tilde\delta (\bold{k_1})
\tilde\delta (\bold{k_2})
\tilde v^i (\bold{k_3})
\tilde v^j (\bold{k_4}) 
\right>.
\eea

\subsection{Linear Regime}

In the linear regime, we consider only the first order terms of $\tilde{\delta}$'s and $\tilde v$'s in $\left< \delta\delta vv\right>$.
Then, Gaussianity of linear $v$ and $\delta$ fields yields
\bea \label{eq:ABCD}
&&\left<
\tilde\delta (\bold{k_1}-\bold{k^\prime})
\tilde v^i (\bold{k^\prime})
\tilde\delta (\bold{k_2}-\bold{k^{\prime\prime}})
\tilde v^j (\bold{k^{\prime\prime}}) 
\right> 
\nonumber\\
&=&
\left<
\tilde\delta (\bold{k_1}-\bold{k^\prime})
\tilde v^i (\bold{k^\prime})
\right> 
\left<
\tilde\delta (\bold{k_2}-\bold{k^{\prime\prime}})
\tilde v^j (\bold{k^{\prime\prime}}) 
\right> 
\nonumber\\
&+&
\left<
\tilde\delta (\bold{k_1}-\bold{k^\prime})
\tilde\delta (\bold{k_2}-\bold{k^{\prime\prime}})
\right> 
\left<
\tilde v^i (\bold{k^\prime})
\tilde v^j (\bold{k^{\prime\prime}}) 
\right> 
\nonumber\\
&+&
\left<
\tilde\delta (\bold{k_1}-\bold{k^\prime})
\tilde v^j (\bold{k^{\prime\prime}}) 
\right> 
\left<
\tilde\delta (\bold{k_2}-\bold{k^{\prime\prime}})
\tilde v^i (\bold{k^\prime})
\right>.
\eea
The first term in the above vanishes and the other terms lead to  \citep{ma/fry:2002}
\bea \label{eq:MF}
&&P_{q_{\perp}}(k,z) = 
\int \frac{d^3k^\prime}{(2\pi)^3} (1-{\mu^\prime}^2) 
\left[
\frac{1}{{k^\prime}^2}
P_{\delta\delta}(|\bold{k}-\bold{k^\prime}|)
P_{\theta\theta} (k^\prime) 
\right.
\nonumber \\ 
&&\left. \qquad \qquad \qquad \qquad 
-\frac{1}{|\bold{k}-\bold{k^\prime}|^2}
P_{\delta \theta}(|\bold{k}-\bold{k^\prime}|)
P_{\delta \theta} (k^\prime) 
\right],
\eea 
where $\theta \equiv \nabla \cdot \bold{v}$. 

The linear velocity is related to the linear density by 
$ \tilde{\theta} = \dot{a} f \tilde{\delta}$, where $f\equiv
d\ln\delta/d\ln a$ and $a(t)$ is the Robertson-Walker scale factor. This
gives the lowest order (LO) expression as \citep{vishniac:1987}
\bea  \label{eq:P_OV}
P^{\rm{LO}}_{q_{\perp}}(k,z) &=& \dot{a}^2f^2\int
 \frac{d^3k^\prime}{(2\pi)^3} P^{lin}_{\delta\delta} (|\bold{k} -
 \bold{k^\prime}|,z) P^{lin}_{\delta\delta}  (k^\prime,z) 
 \nonumber \\
& &\qquad\qquad\times
\frac{k(k - 2k^\prime\mu^\prime)(1-{\mu^\prime}^2)}{{k^\prime}^2(k^2 +
{k^\prime}^2-2kk^\prime\mu^\prime)},
\eea
where $P^{lin}_{\delta\delta} $ is the linear matter density power
spectrum. A similar derivation for the power spectrum of the sum of
longitudinal and transverse momentum fields is presented in
\cite{2000MNRAS.319..573P}. 

Taking $k\rightarrow 0$ limit of Equation~(\ref{eq:P_OV}), we find
\bea \label{eq:P_OV^{lowk}}
P^{\rm{LO}}_{q_{\perp}}(k,z) \to \frac{8}{15}\dot{a}^2f^2k^2
\int  \frac{dk^\prime}{(2\pi)^2} \frac{{[P^{lin}_{\delta\delta}  (k^\prime,z)]^2 }}{{k^\prime}^2}.
\eea
The $k$ dependence of $P^{\rm{LO}}_{q_{\perp}}$ is thus given simply by
$k^2$, which is independent of cosmology or the initial power spectrum. 
Similarly, the second order term (2nd term in the r.h.s. of Eq.~\ref{eq:P_q_par}) of the longitudinal mode also goes as $k^2$ in the low-$k$ limit:
\bea
P^{\rm{(2)}}_{q_{\parallel}}(k,z) \to \frac{7}{15}\dot{a}^2f^2k^2
\int  \frac{dk^\prime}{(2\pi)^2} \frac{{[P^{lin}_{\delta\delta}  (k^\prime,z)]^2 }}{{k^\prime}^2}.
\eea
We shall discuss the physical origin of this dependence in
Section~\ref{sec:connected_moment}. 

We use CAMB to compute $P_{\delta\delta}^{lin}$. The dashed line in
Figure~\ref{fig:P_q_sim} shows $P^{\rm{LO}}_{q_{\perp}}$ at $z=0$, while the
dotted line shows the low-$k$ limit (Eq.~\ref{eq:P_OV^{lowk}}).

\begin{figure*}  
  \begin{center} 
      \includegraphics[scale=0.55]{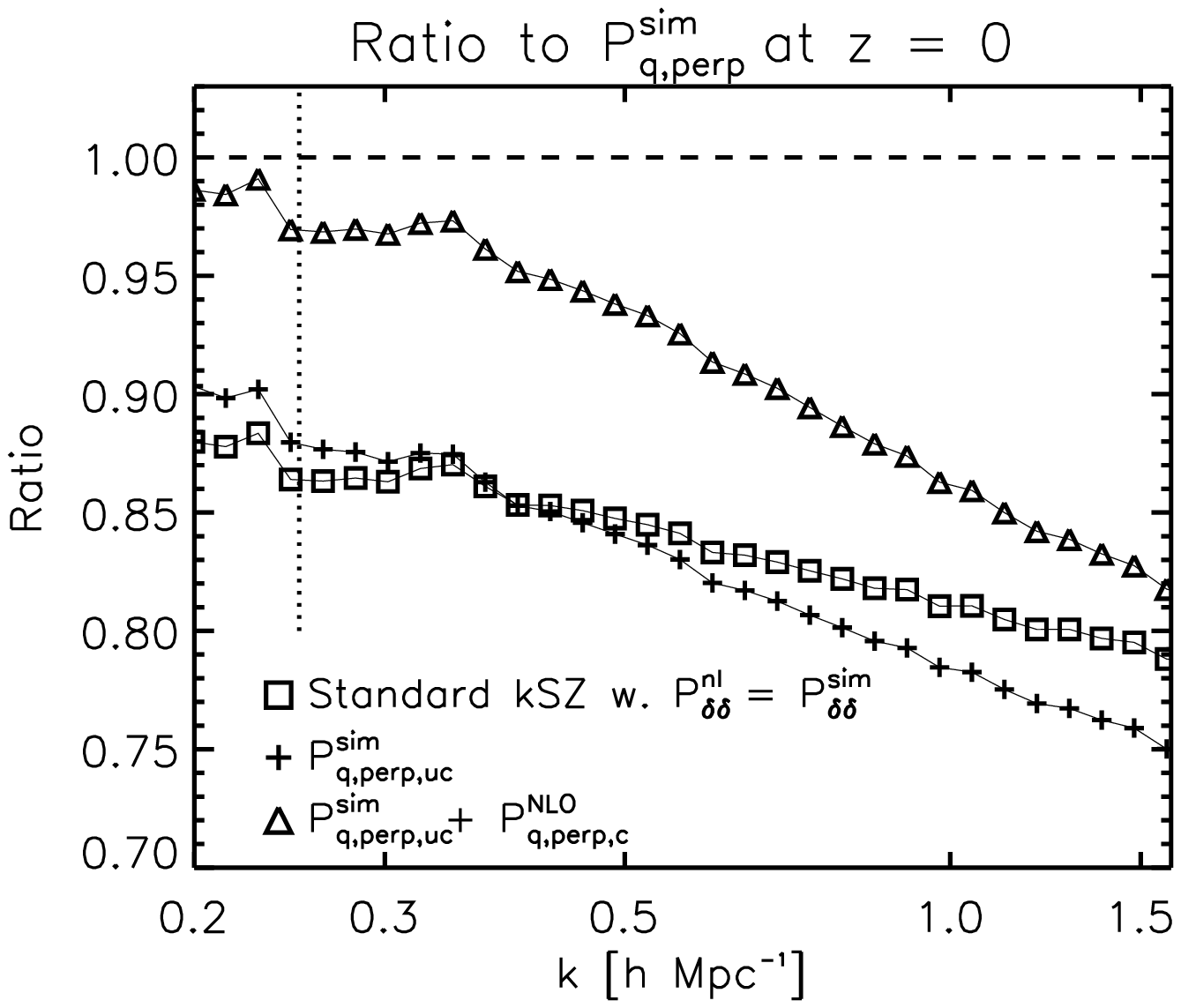}
      \includegraphics[scale=0.55]{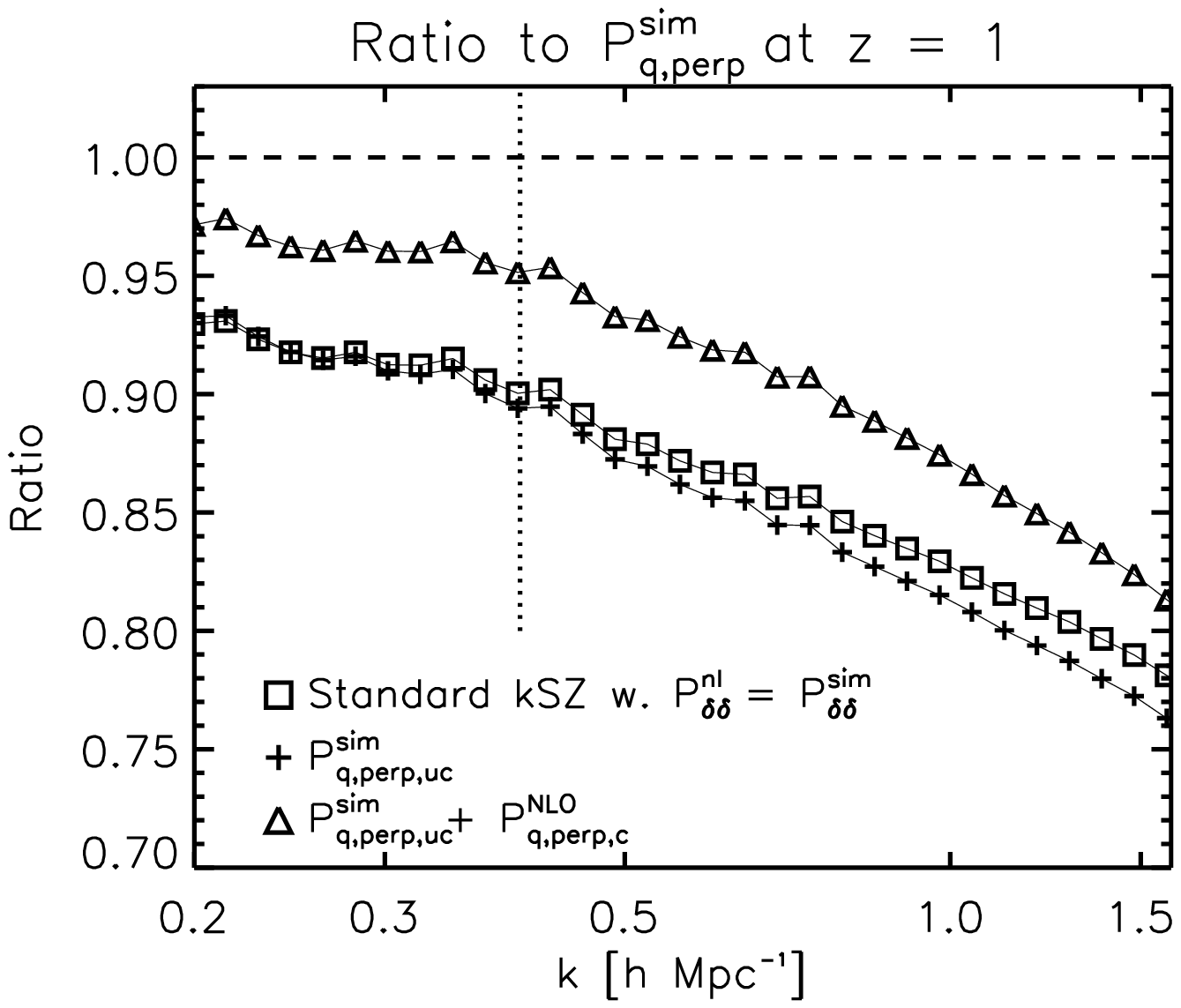}
      \includegraphics[scale=0.55]{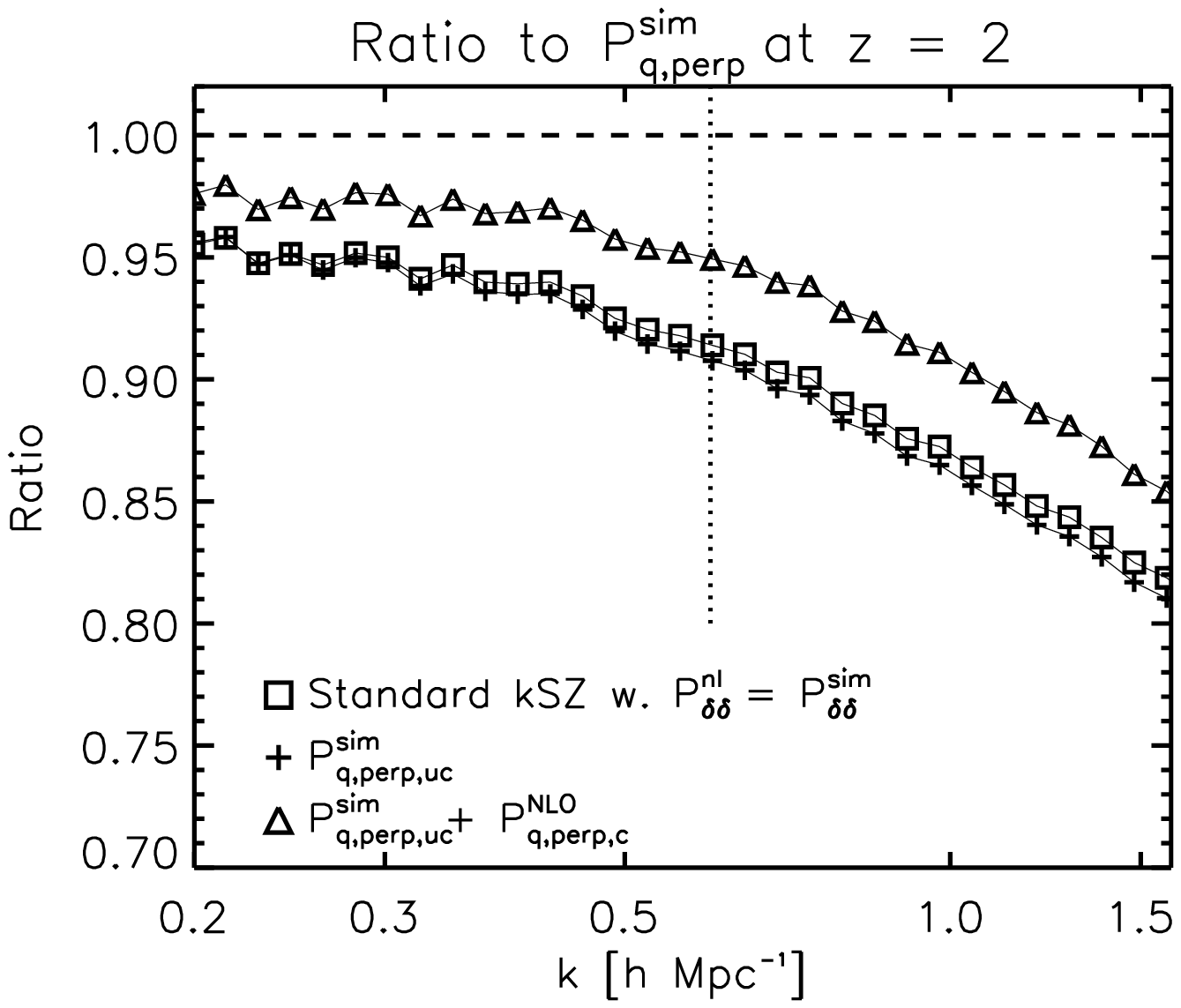}
      \includegraphics[scale=0.55]{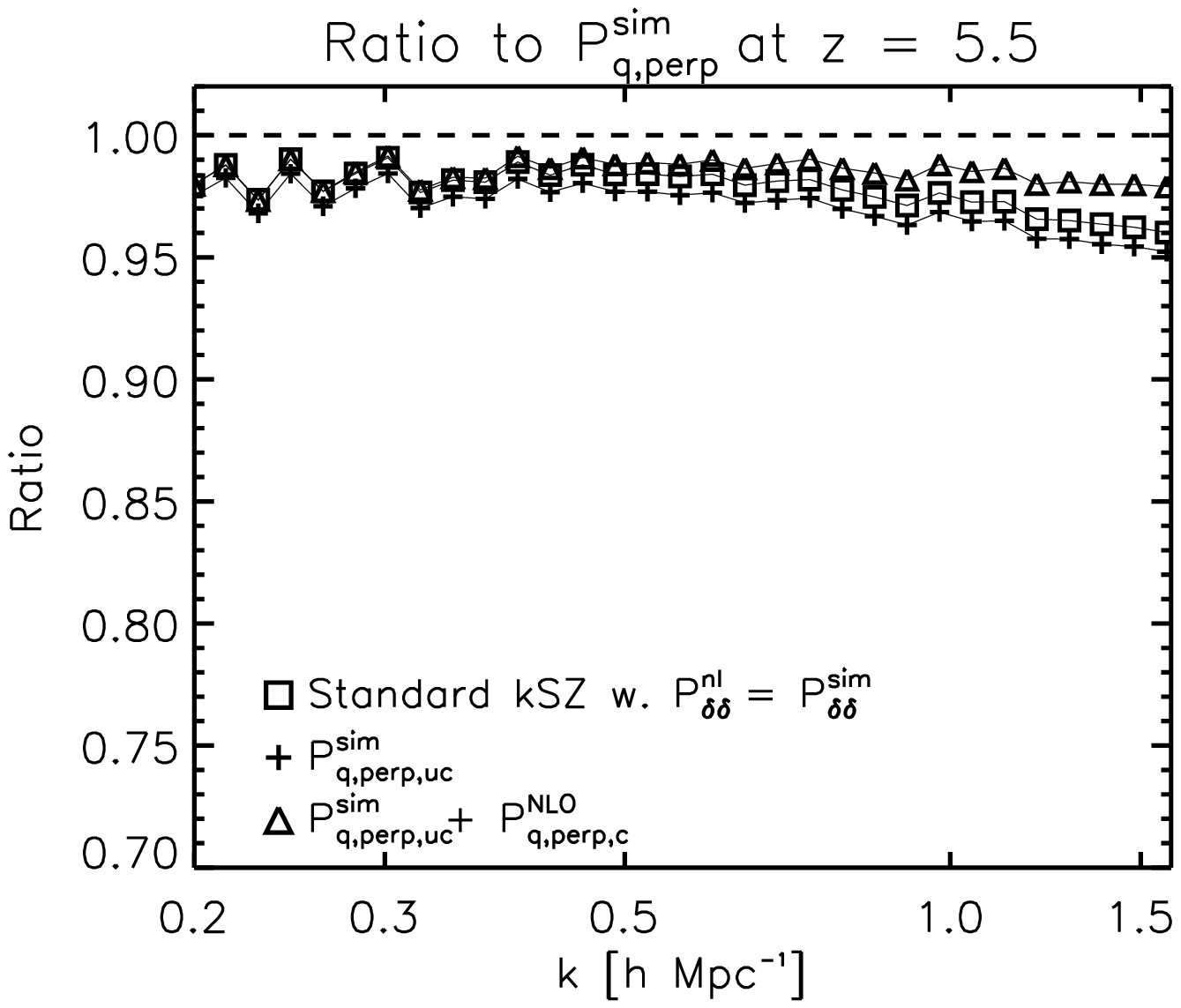}
        \caption{Ratios of various models for $P_{q_\perp}$ to
   $P_{q_\perp}$ directly measured from the simulation at $z=0$ (upper
   left), $z=1$ (upper right), $z=2$ (lower left) and $z=5.5$ (lower
   right). The models include the Standard expression (Eq.~\ref{NLOV}) with
   $P^{nl}_{\delta\delta}$ from the simulation (squares),
   the unconnected term (Eq.~\ref{eq:MF}) with 
   $P_{\delta\delta}$, $P_{\delta\theta}$ and $P_{\theta\theta}$ from the simulation (crosses), 
   and the connected term calculated from perturbation theory added to the crosses (triangles).
   The vertical dotted line denotes the wavenumber beyond which the perturbation theory fails to model nonlinearity precisely according to \cite{2006ApJ...651..619J} and our data. At $z=5.5$, that wavenumber is out of the plotting range and not shown in the figure.
   }
        \label{fig:test_kSZ_model}
  \end{center}  
\end{figure*}

\subsection{Nonlinear Regime}\label{sec:NL}

As Figure~\ref{fig:P_q_sim} shows, nonlinear contributions to
$P_{q_\perp}$ become dominant over the LO contribution at  $k\gtrsim
0.2~h/\rm{Mpc}$ at $z=0$, in agreement with the previous work
\citep{hu:2000,ma/fry:2002}. The current popular model of the
post-reionization kSZ signal uses an approximate expression for the
transverse momentum power spectrum  due to
\cite{hu:2000}, which replaces one of $P_{\delta\delta}^{lin}$ in
Equation~(\ref{eq:P_OV}) with the nonlinear power spectrum,  $P_{\delta\delta}^{nl}$:
\begin{eqnarray} \label{NLOV}
\nonumber 
P^{\rm{S}}_{q_{\perp}}(k,z) &=& \dot{a}^2f^2\int \frac{d^3k^\prime}{(2\pi)^3} P_{\delta\delta}^{nl} (|\bold{k} - \bold{k^\prime}|,z) P^{lin}_{\delta\delta}(k^\prime,z)\\
&&\qquad \qquad\times
\frac{k(k - 2k^\prime\mu^\prime)(1-{\mu^\prime}^2)}{{k^\prime}^2(k^2 + {k^\prime}^2-2kk^\prime\mu^\prime)}.
\end{eqnarray}
This model modifies the second term in the square bracket of Equation~(\ref{eq:MF}) in the following way:
\bea \label{hyp1}
P_{\delta \theta}(|\bold{k}-\bold{k}^\prime|)P_{\delta \theta}(k^\prime) 
= P_{\delta \delta}(|\bold{k}-\bold{k}^\prime|)P_{\theta \theta}(k^\prime). 
\eea 
This holds in the linear regime, and the entire term vanishes in the
large $k$ regime due to the pre-factor,
$|\bold{k}-\bold{k^\prime}|^{-2}$. In addition, the model
approximates the velocity power spectrum by linear theory, i.e.,
$P_{\theta \theta}=\dot{a}^2 f^2 P^{lin}_{\delta\delta}$. In this way, the model
avoids having to model the velocity power spectrum and the
density-velocity cross power spectrum, which is relatively poorly
understood. We shall refer to this model as the {\it Standard}
model (hence the superscript ``S'' in Eq.~\ref{NLOV}) in this paper.

However, Equation~(\ref{eq:MF}), which the Standard model aims to model,
is not the full expression for $P_{q_\perp}$ because it neglects the
contribution from the connected term, $\langle\delta v\delta
v\rangle_c$. In the nonlinear regime, nonlinear growth makes both
$\delta$ and $v$ non-Gaussian, and thus there is no reason to think that
the connected four-point term is negligible. We shall quantify the importance of this term in detail in Section~\ref{sec:connected_moment}.

%\section{Revisiting the Standard Model for the Effect of Nonlinearity on the Power Spectrum of Transverse Momentum Fluctuations}\label{sec:results}
\section{Revisiting the Standard Model for the Effect of Nonlinearity on the Transverse Momentum Power Spectrum}\label{sec:results}

\subsection{Unconnected Term} \label{sec:test_kSZ}

In this section, we revisit accuracy of Equation~(\ref{NLOV}). As noted above, this model is an approximation for the
unconnected term (Eq.~\ref{eq:MF}). Therefore, it, by design, does not
take into account the connected term. 

We test whether Equation~(\ref{NLOV}) successfully approximates
Equation~(\ref{eq:MF}) by evaluating it using the density power spectrum
from the simulation, i.e., $P^{nl}_{\delta\delta}=P^{sim}_{\delta\delta}$, and
compare it with Equation~(\ref{eq:MF}) using $P_{\delta\delta}$,
$P_{\delta \theta}$, and $P_{\theta\theta}$ from the simulation. 

In Figure~\ref{fig:test_kSZ_model}, we show the ratios of
Equation~(\ref{NLOV}) (squares) and Equation~(\ref{eq:MF}) (crosses) to
$P_{q_\perp}$ measured directly from the simulation. At $0.2\lesssim k
\lesssim 0.5~h~\rm{Mpc}^{-1}$ at $z=0$ (top left panel), the Standard
model reproduces the unconnected term with high accuracy ($\sim2\%$ level). However, the
Standard model overestimates the unconnected term at larger $k$, reaching $\sim5\%$ level at $k\sim 1~h~\rm{Mpc}^{-1}$. We attribute this error to the linear velocity assumption overestimating magnitudes of velocity modes. In the high-$k$ limit, Equation~(\ref{eq:MF}) converges to \citep{hu:2000}
\bea
P_{q_\perp}(k) = \frac{2}{3}P_{\delta\delta}(k) v_{rms}^2(k),
\eea
where $v_{rms}^2(k)=\int_{k^\prime\le k} \frac{d^3k^\prime}{(2\pi)^3}
P_{vv}(k^\prime)$ is the velocity dispersion. Nonlinear correction
makes the velocity power spectrum in the relevant $k$ range {\it
smaller} than the linear velocity power spectrum \citep[see, e.g.,
Figure 1 of][]{pueblas/scoccimarro:2009}; thus, linear theory
overestimates $v_{rms}^2$. This effect becomes smaller at higher
redshifts, as shown in the other panels of Figure~\ref{fig:test_kSZ_model}.

We find that both Equation~(\ref{eq:MF}) and (\ref{NLOV}) underestimate
$P_{q_\perp}$ significantly compared to $P_{q_\perp}$ measured directly
from the simulation. The underestimation decreases monotonically with
redshift: $12-25$\% over $k=0.2-1.5~h~{\rm Mpc}^{-1}$ at $z=0$; $7-23$\% at
$z=1$; $5-18$\% at $z=2$; and a few to 5\% at $z=5.5$. Thus, this is
likely related to the development of nonlinear structure formation.

\subsection{Connected Term} \label{sec:connected_moment}
	
\begin{figure}
  \begin{center}
    \includegraphics[scale=0.63]{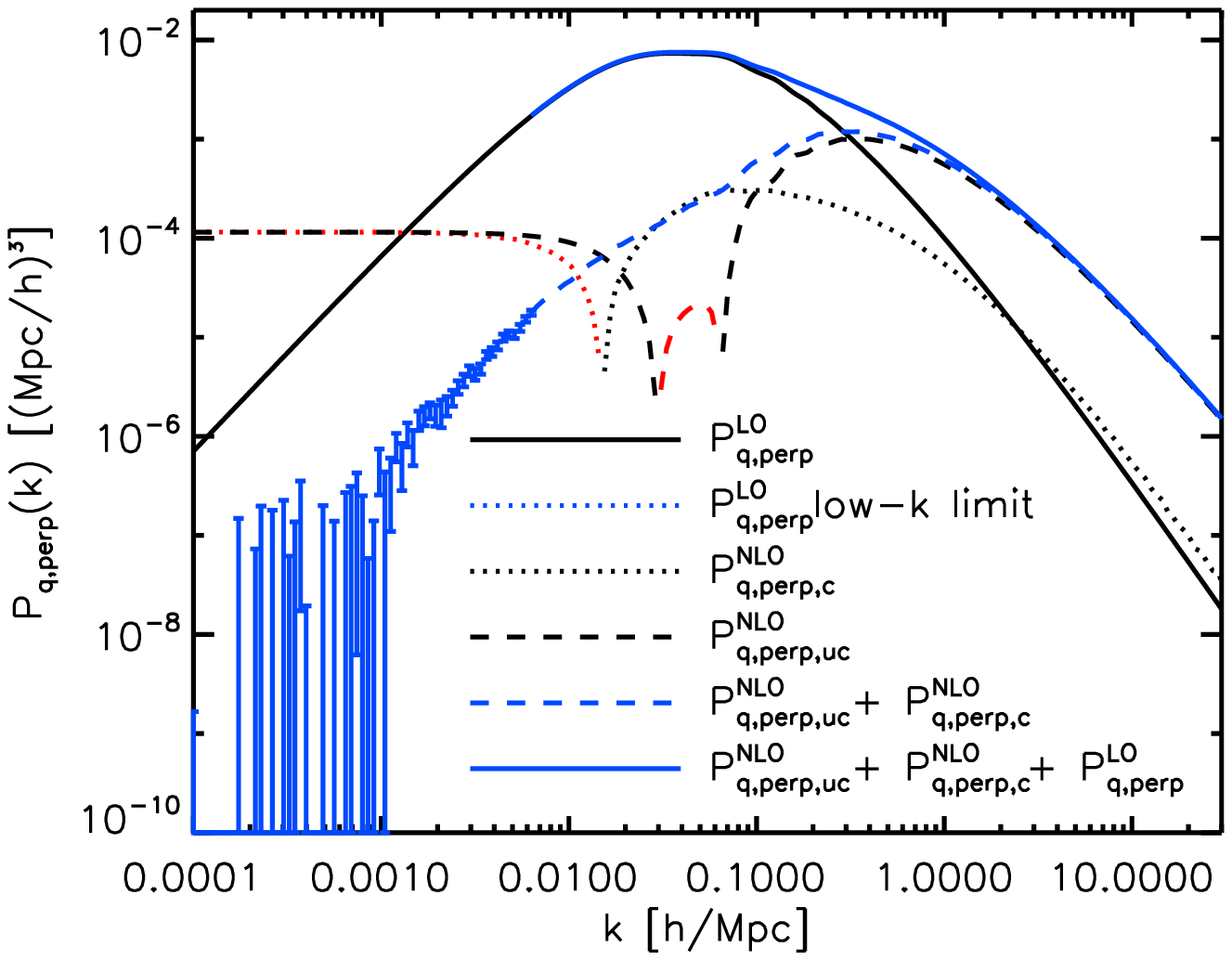}
  \caption{$P_{q_\perp}$ at $z=0$ from the next-to-leading order
   unconnected term (black dashed), connected term (black dotted), the sum of
   the two next-to-leading order terms (blue dashed), and the total sum of the leading and next-to-leading order terms (blue solid). The red color shows negative values. The error
   bars show the uncertainty of the Monte-Carlo integration for $k<0.01~h~\rm{Mpc}^{-1}$. The solid line shows the LO term
   (Eq.~\ref{eq:P_OV}), while the blue dotted line shows its
   low-$k$ limit (Eq.~\ref{eq:P_OV^{lowk}}).}
    \label{fig:P_q_c}
  \end{center}
\end{figure}

To show that the connected term is the likely explanation for the
difference between the simulation data and the models that ignore the
connected term, we calculate the connected term using perturbation
theory. Since the connected term vanishes in linear theory, we must go
to the next-to-leading order perturbation theory, such as the standard
``one-loop'' perturbation theory \citep{2002PhR...367....1B}. This
theory allows us to extend validity of analytic solutions for Fourier
modes of density and velocity fields down to weakly nonlinear scales,
e.g., in $k \lesssim 0.3~h~\rm{Mpc}^{-1}$ at $z\sim 1$ and in wider wavenumbers
at higher redshifts \citep{2006ApJ...651..619J}. 
We also confirm that the density power spectrum from our simulation data supports their findings.
We derive the explicit expressions for the connected (Eq.~\ref{eq:P_q,c}) and unconnected terms
(Eq.~\ref{eq:P_q,uc}) in Appendix~\ref{sec:P_q,c} and \ref{sec:P_q,uc},
and show them in the dotted and dashed lines in Figure~\ref{fig:P_q_c},
respectively.

We find that each of the unconnected and connected terms does not
vanish in the low-$k$ limit on its own. Instead, they converge to a
constant value with opposite signs. This constant can be obtained by
taking $k\to 0$ limit of Equation~(\ref{eq:P_q,c}):
\bea
P_{q_\perp,c}|_{k\rightarrow 0}
&=&
\frac{16}{49}
\int
\frac{d^3k^\prime}{(2\pi)^3}
\int
\frac{d^3k^{\prime\prime}}{(2\pi)^3}
\frac{[ 1 - (\hat\bold{k} \cdot \hat\bold{k^{\prime}}) (\hat\bold{k} \cdot \hat\bold{k^{\prime\prime}}) ]}
{k^\prime k^{\prime\prime}}
\nonumber \\ 
&\times&P(\bold{k^\prime})
P(  \bold{k^\prime} + \bold{k^{\prime\prime}})
P(\bold{k^{\prime\prime}} )
\left(
\frac{ %%%
(1 - \mu^2){k^{\prime\prime}}{k^{\prime}}
} %%%
{ %%%
(\bold{k^\prime}+\bold{k^{\prime\prime})}^2
} \right)^2.~~~~~%%%
\eea
The sum of the two terms, however, yields a physical result that remains
positive and decays toward lower $k$. The cancellation of two large numbers
introduces an uncertainty in our numerical integration using the Monte
Carlo method\footnote{In principle we can derive the low-$k$ limit of
the next-to-leading order expression analytically. Here, we have chosen
to perform numerical integration because of the complexity of the
results given in Equations~(\ref{eq:P_q,c}) and (\ref{eq:P_q,uc}).}. Within the uncertainty, the result is consistent with
$k^2$ scaling, which is the same as the scaling of the LO
expression given by Equation~(\ref{eq:P_OV^{lowk}}). This calls for a
physical explanation; namely, what is the physical origin of the $k^2$
scaling in the low-$k$ limit of the transverse momentum power spectrum,
which is independent of cosmology or the initial power spectrum?

\begin{figure*}
  \begin{center}
    \includegraphics[scale=0.75]{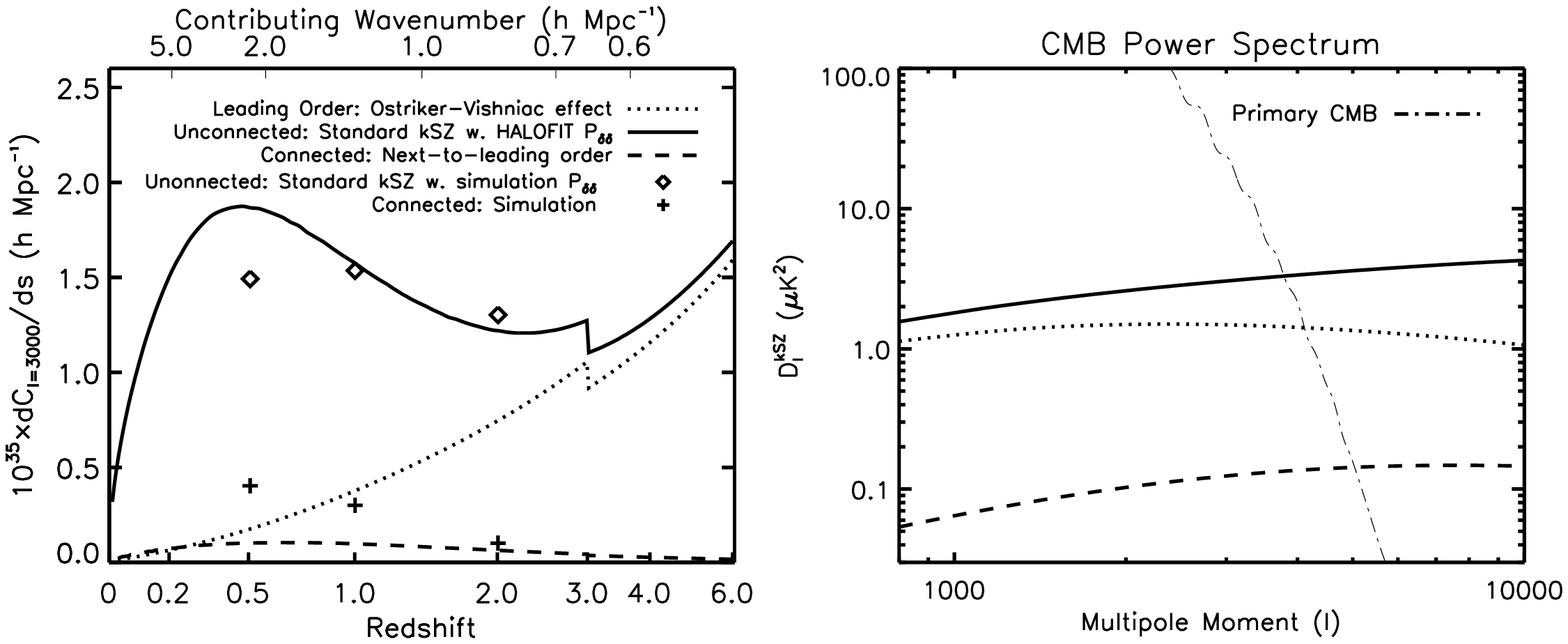}
  \caption{Angular power spectrum of the kSZ effect. (Left)
   $dC_{\ell=3000}/ds$, which shows the contribution to $C_\ell$ at $\ell=3000$
   from a given comoving distance. The horizontal
   axis shows redshifts such that it is linear in the comoving
   distance. The top label shows the contributing wavenumber to the kSZ
   signal, $\ell/s(z)$. The discontinuity at $z=3$ is due to the
   instantaneous helium reionization. The solid line shows the Standard
   kSZ model (Eq.~\ref{NLOV}) with the HALOFIT density power spectrum,
   the dashed line shows the next-to-leading order connected
   term from perturbation theory, and the dotted line shows the LO
   contribution (Eq.~\ref{eq:P_OV}). We also show the simulation results
   at $z=0.5$, 1, and 2: the diamonds show Equation~(\ref{NLOV}) with the
   nonlinear density power spectrum from the simulation, while the
   crosses use the total 
   simulation momentum power spectrum minus the diamonds. 
  (Right) $D_\ell=\ell(\ell+1)C_\ell/(2\pi)$. The dash-dotted line
   shows the primary CMB temperature power spectrum, while the other lines
   show the same cases as in the left panel.}
  \label{fig:integrand}
  \end{center}
\end{figure*}

As shown by \cite{mercolli/pajer:2014}, this is a consequence of
momentum conservation. In short, the argument goes as follows. Suppose that we have a uniform distribution of
matter particles with no initial momentum or density fluctuation, and
displace these particles with velocities $\bold{v}_i$ (where $i$ is a
particle ID) in a momentum-conserving manner. In this way, we have
removed the effect of the initial condition, and can focus only on the
effect of the subsequent evolution of particle's motion. Fourier transform of
momentum of the displaced particles is given by
$\bold{q}(\bold{k})=\sum_im_i\bold{v}_i\exp(-i\bold{k}\cdot x_i)$, and the
low-$k$ limit is
$\bold{q}(\bold{k})=\sum_im_i\bold{v}_i(1-i\bold{k}\cdot\bold{x}_i+\dots)$. The
first term vanishes by momentum conservation, giving
$\bold{q}(\bold{k})={\cal O}(k)$; hence the power spectrum of momentum
fields is proportional to $k^2$. This argument applies to both the
longitudinal and transverse momentum fields. As we have found, the
connected term is necessary for obtaining the correct low-$k$ limit at
the next-to-leading order in perturbation theory.

\section{Implication for the Post-reionization kSZ angular power spectrum}\label{sec:PRKSZ}

In the high-$k$ regime, the connected term is of order ten percent of
the unconnected term, which brings the predicted $P_{q_\perp}$ into
better agreement with the simulation, as shown by the triangles in
Figure~\ref{fig:test_kSZ_model}. Thus, the underestimation is now much
reduced to $2-18$\% over $k=0.2-1.5~h~{\rm Mpc}^{-1}$ at $z=0$; $2-18$\%
at $z=1$; $2-14$\% at $z=2$; and a few percent at $z=5.5$. 
For each redshift, we mark in the figure roughly the wavenumber that the next-to-leading order (``one-loop'') calculation of perturbation
theory becomes inaccurate for the density fluctuations according to \cite{2006ApJ...651..619J} and our data.
The remaining differences beyond that wavenumber are likely due to inaccuracy
of perturbation theory calculation, whereas the agreement at 2\% level at small $k$ supports our
conclusion that the connected term is necessary for accurate modeling of
the transverse momentum power spectrum.

We use Equation~(\ref{C_l}) to calculate the observable CMB angular
power spectrum of kSZ, $C_\ell$. In the left panel of
Figure~\ref{fig:integrand}, 
we show $dC_{\ell=3000}/ds$, which is
the contribution to the kSZ signal at $\ell=3000$ per comoving
distance from the observer. The area under each
curve gives the total $C_\ell$, which is shown in the right panel. The
dotted line shows the LO calculation (Eq.~\ref{eq:P_OV}). 
The solid line and the diamonds show the Standard kSZ model
(Eq.~\ref{NLOV}) with the nonlinear matter density power spectrum
computed using the HALOFIT code \citep{smith/etal:2003} and the
simulation (i.e., the square symbols in
Figure~\ref{fig:test_kSZ_model}), respectively. We cannot calculate the
kSZ contribution at $z<0.5$ in our simulation because the contributing
wavenumber is beyond the resolution limit of our simulation ($k\sim
2~h~\rm{Mpc}^{-1}$). The disagreement between the two at $z=0.5$ is due
to the disagreement between the HALOFIT power spectrum and our
simulation at $k\gtrsim1~h~\rm{Mpc}^{-1}$, which is also due to the
resolution limit of the simulation.

The dashed line shows the next-to-leading order connected moment
contribution. We also estimate the connected term contribution in our
simulation by subtracting the diamonds from the total
transverse momentum power spectrum measured from the simulation. The
simulation at $z=1$ suggests about a factor of 3 larger effect than the next-to-leading order perturbation theory. The
interpretation of the difference at $z=0.5$ is difficult because of the
disagreement between the diamond and the solid line; however, if we
assume that the resolution of the simulation affects the unconnected and
connected terms in the same way, then the simulation at $z=0.5$ suggests
about a factor of three larger effect than perturbation theory. This is
expected because the kSZ signal receives contributions from $k\gtrsim
1~h~\rm{Mpc}^{-1}$ where the next-to-leading order calculation is
insufficient to capture nonlinearity (see the triangles in 
Figure~\ref{fig:test_kSZ_model}). The difference
decreases as $z$ increases, as expected from development of nonlinear
structure formation. Taking into account this extra nonlinearity above
perturbation theory, we estimate that the connected term contribution to
$C_\ell$ at $\ell=3000$ is at least ten percent of the unconnected
term. 

%The solid line in the left panel of Figure~\ref{fig:integrand} illustrates how the kSZ signal accumulates as the function of the distance from observer for $\ell=3000$ in the current model of nonlinear transverse momentum power spectrum (For details of the model, see Section~\ref{sec:basics}). For comparison, the dotted line shows the hypothetical case that the density field and the velocity field are in the linear regime. Comparison of these two cases highlights the importance of the nonlinearity at $z=3$. The resulting kSZ power spectrum, $D^{\rm{kSZ}}_\ell$, in the right panel (same line types for linear and nonlinear regimes as in the left panel) shows that nonlinear enhances the total signal by a factor of two or greater.

% For the integrated signal, $C_\ell$, the next-to-leading order connected moment amounts to $\sim 4\%$ of the unconnected moment. Although its relative magnitude in the transverse momentum power spectrum is larger ($\sim 10\%$) at $z=0$, it decays as $\sim \dot{a}^2f^2 \delta^6$ as the redshift increases. The leading order part of the unconnected moment is decays more slowly as $\sim\dot{a}^2f^2 \delta^4$ making the connected moment subdominant at $z\gtrsim 1$. Thus, the connected moment contribution in the total $C_\ell$ is diluted compared to its significance at $z=0$. 

\section{Summary and Conclusion} \label{sec:Conclusion}
We have reexamined the currently popular model of the transverse dark
matter momentum power spectrum (Eq.~\ref{NLOV}) used in the previous
calculation of the post-reionization kSZ power spectrum
\citep[e.g.,][]{shaw/etal:2011}. We find that the current model
reproduces the contribution of the unconnected term (Eq.~\ref{eq:MF})
well. However, this model ignores the contribution from the connected
term that arises in the nonlinear regime. Using both perturbation
theory and cosmological $N$-body simulation, we show that the
contribution from the connected term adds a significant additional
power, especially  at larger $k$ at lower $z$
(Figure~\ref{fig:test_kSZ_model}). This is consistent with the
expectation from nonlinear structure formation. 

%$k\gtrsim 0.1~h~\rm{Mpc}^{-1}$ at $z=0$. We derive its expression in the next-to-leading order (Eq.~\ref{eq:P_q,c}), which explains the difference in power spectrum between the current kSZ model alone and our simulation result at quasi-nonlinear scale ($0.1\lesssim k \lesssim 1~h~\rm{Mpc}^{-1}$). At higher wavenumbers ($k \gtrsim 1~h~\rm{Mpc}^{-1}$), the connected moment contribution from the simulation is higher than what the next-to-leading order expression gives due to additional contribution from higher order terms.

%In the angular power spectrum of CMB temperature fluctuations that the
%kSZ effect induces, the next-to-leading order connected moment
%contribution amount to only $\sim 4 \%$ of what the unconnected moment
%contribution amounts to at $\ell=3000$. But, higher ($\sim 10\%$) order
%contribution from our simulation suggests a larger contribution. 

We estimate that the contribution of the connected term to the kSZ
angular power spectrum at $\ell=3000$ is ten percent of the unconnected
term. This is the term that needs to be added to the calculation of
\cite{shaw/etal:2011}, assuming that a similar correction is necessary
for the momentum power spectrum of gas. In light of the current
observational constraint ($D^{\rm kSZ}_{\ell=3000}=2.9~\mu K^2$) and the
post-reionization kSZ model of \cite{shaw/etal:2011} ($D^{{\rm
kSZ},z<6}_{\ell=3000}=2.0~\mu K^2$), adding ten percent to the
post-reionization kSZ signal would imply a twenty percent less kSZ signal
from the EoR. The semi-numerical model of \citet{battaglia/etal:2012b}
implies that such a change in the reionization kSZ signal would result in
a twenty percent increase in the redshift at which the mean ionized fraction of the
universe reached 50\%, or ten percent
shorter duration of reionization, $\Delta z$. 
\cite{2013ApJ...769...93P}, however, show that reionization simulations
based upon radiative transfer and $N$-body simulation can yield
a more extended duration of the reionization epoch than these
semi-numerical models predict, so this quantitative conclusion
about the constraint from changing the kSZ upper limit from the
EOR on the duration of reionization may need to be
revised to take account of such extended reionization.

Finally, we have shown that both the LO and next-to-leading order
perturbation theory calculations give $P_{q_\perp}\propto k^2$ in the
low-$k$ limit, independent of cosmology or the initial power
spectrum. This behavior is consistent with momentum conservation.

%Lastly, we note 
%that difference in the distribution of the dark matter and the baryonic matter is also not reliable to $\sim 10\%$ is the signal at $k\gtrsim 1~h~\rm{Mpc}^{-1}$. In order to constrain the post-reionization kSZ to the level meaning for reionization study, high resolution-

%We also note our limitation in covering all the wavenumbers that contribute to $dC_{\ell=3000}$. $dC_{\ell=3000}/ds$ in the left panel of Figure~\ref{fig:integrand} suggests that covering more than 90\% over the kSZ signal would require us to resolve the momentum field down to $k\sim 5~h~\rm{Mpc}^{-1}$. This is beyond the wavenumber coverage of our simulation and this is barely feasible with the current computational power. While it is clearly not feasible to run such simulation in a volume same as ours, it is suggested a smaller simulation with the size of the box down to $100~h^{-1}~\rm{Mpc}$ can reliably corrected for its the missing velocity mode using linear approximation. Also, baryonic physics has to be taken into account carefully as the distribution of the gas deviate significantly from that of the dark matter at such small scales. Although the current model attempts to account for this, it is not yet reliable down to $\sim10\%$ level. Therefore, high-resolution $N$-body hydrodynamics simulations are needed to accurately subtract the post-reionization kSZ to improve constraining power for the epoch of reionization. We leave this to future studies.
\section{Acknowledgment} 
Authors thank Ilian T. Iliev for generating and storing the cosmological $N$-body simulation data used in this work. 
PRS was supported in part by U.S. NSF grant AST-1009799, NASA grant NNX11AE09G, NASA/JPL grant RSA Nos. 1492788 and 1515294, and supercomputer resources from NSF XSEDE grant TG-AST090005 and the Texas Advanced Computing Center (TACC) at the University of Texas at Austin. JK was supported by the Australian Research Council Centre of Excellence for All-sky Astrophysics (CAASTRO), through project number CE110001020. YM was supported by the Labex ILP (reference ANR- 10-LABX-63) part of the Idex SUPER, and received financial state aid managed by the Agence Nationale de la Recherche, as part of the programme Investissements d'avenir under the reference ANR-11-IDEX-0004-02. 

\appendix

\section{kSZ effect from Longitudinal Modes} \label{sec:P_q,par}
Rewriting Equation~(\ref{Eq:kSZ2}) in terms of Fourier mode of $\bold{q}$ gives
\bea
\frac{\Delta T}{T} (\hat{\gamma}) = -\frac{\sigma_T n_{e,0}}{c}  \int \frac{ds}{a(s)^2} e^{-\tau} \int \frac{d^3k}{(2\pi)^3} [\hat{\gamma} \cdot \tilde{\bold{q}}(\bold{k},s)] e^{-i\bold{k}\cdot(s\hat{\gamma})}.
\eea
Here, the momentum vector in Fourier space, $\tilde{\bold{q}}$, can be decomposed into the longitudinal mode, $\tilde{q}_\parallel \equiv \tilde{\bold{q}}\cdot \hat{k}$, and the transverse component, $\tilde{q}_\perp\equiv |\tilde{\bold{q}} -\hat{k} (\tilde{\bold{q}} \cdot \hat{k}) |$, to give
\bea
\frac{\Delta T}{T } (\gamma) = -\frac{\sigma_T n_{e,0}}{c}  \int \frac{ds}{a(s)^2} e^{-\tau} \int \frac{d^3k}{(2\pi)^3}
[x\tilde{q}_\parallel + \cos(\phi_{\hat{q}} -\phi_{\hat{\gamma}}) (1-x^2)^{1/2}\tilde{q}_\perp (\bold{k},s)]e^{-iksx},
\eea
where $x\equiv\hat{k} \cdot \hat{\gamma}$. 

In this section, we shall derive the angular power spectrum of $q_\parallel$ term.
The derivation given here parallels that for the transverse mode given in Appendix of \cite{2013ApJ...769...93P}.
Spherical-harmonics decomposition of $q_\parallel$ term in the above equation is,
\bea
a_{\ell m} &=&  \int d^2 \hat{\gamma} 
Y^{m*}_\ell (\hat{\gamma}) 
\frac{\Delta T}{T} (\hat{\gamma}) 
\nonumber\\
&=& 
- \frac{\sigma_T n_{e,0}}{c} 
\int d^2\hat\gamma 
Y^{m*}_\ell (\hat{\gamma}) 
\int \frac{ds}{a(s)^2} e^{-\tau}
\int \frac{d^3k}{(2\pi)^3} 
x \tilde{q_\parallel} 
(\bold{k},s)
e^{-iksx}
\nonumber\\
&=&
\int \frac{d^3k}{(2\pi)^3} 
f_{\ell m} (\bold{k})
\eea
where
\bea
f_{\ell m}(\bold{k}) 
&\equiv &
\int d^2 \hat{\gamma} Y^{m*}_\ell (\hat{\gamma})
\int \frac{ds}{a(s)^2} e^{-\tau} x \tilde{q}_\parallel (\bold{k},s) e^{-iksx}
\nonumber\\
&=&
\int d^2 \hat{\gamma} Y^{m*}_\ell (\hat{\gamma})
\int \frac{ds}{a(s)^2} e^{-\tau} x \tilde{q}_\parallel (\bold{k},s) \times 
4\pi
\sum_{LM} 
(-i)^L
j_L (ks)
Y^M_L (\hat{\gamma})
Y^{M*}_L (\hat{k}).
\eea

Here, we choose a coordinate system that $\hat{k} = \hat{z}$, thereby simplifying $Y^{M*}_L(\hat{k})$ to $Y^{M*}_L(\hat{z}) = \delta_{M0} \sqrt{\frac{2L+1}{4 \pi}}$. 
Then, 
\bea
f_{\ell m}(k\hat{z}) &=& \sqrt{4\pi} \int \frac{ds}{a(s)^2} e^{-\tau} \tilde{q}_\parallel (\bold{k},s) \sum_L (-i)^L j_L (ks) \int d^2 \hat{\gamma} Y^0_L (\hat{\gamma}) 
\cos{\theta} Y^{m*}_\ell (\hat{\gamma})
\nonumber \\
&=& 
\sqrt{\frac{16 \pi^2}{3}} \int \frac{ds}{a(s)^2} e^{-\tau} \tilde{q}_\parallel (\bold{k},s) \sum_L (-i)^L j_L (ks) \int d^2 \hat{\gamma} Y^0_L (\hat{\gamma}) 
Y^0_1(\hat{\gamma}) Y^{m*}_\ell (\hat{\gamma})
\eea
where $\theta$ is the azimuthal angle of $\hat{\gamma}$ in the coordinate system of our choice. Also, we have used the fact that $Y^0_1(\theta,\phi)= \frac{1}{2}\sqrt{\frac{3}{\pi}} \cos{\theta}$. 

The integral over $\hat{\gamma}$ can be given in terms of the Clebsch-Gorden coefficient:
\bea
\int d^2 \hat{\gamma} Y^0_L(\hat{\gamma}) Y^0_1(\hat{\gamma}) Y^{m*}_\ell(\hat{\gamma}) = \sqrt{\frac{3(2\ell + 1)}{4\pi (2L+1)}}
C_{\ell 1} (L,0;m,0) C_{\ell 1} (L,0;0,0)\delta_{0,m}.
\eea
In this case, the relevant ones are 
\bea
C_{\ell 1} (\ell+1,0;0,0) = \sqrt{\frac{\ell + 1}{2\ell + 1}}, ~
C_{\ell 1} (\ell,0;0,0) = 0, ~
C_{\ell 1} (\ell-1,0;0,0) = \sqrt{\frac{\ell}{2\ell + 1}},
\eea
and, they give
\bea
f_{\ell,m=0} (k\hat{z}) &=& 
\sqrt{4\pi (2\ell + 1)} \int\frac{ds}{a(s)^2} e^{-\tau} \tilde{q}_\parallel (\bold{k},s) \sum_L (-i)^L j_L (ks) \times 
C_{\ell 1} (L,0;0,0) C_{\ell 1} (L,0;0,0)
\nonumber \\
&=& 
(-i)^{\ell + 1} 
\sqrt{4\pi (2\ell + 1)}
\int\frac{ds}{a(s)^2} e^{-\tau}
\left[
\frac{\ell + 1}{2\ell + 1} j_{\ell + 1} (ks)
-
\frac{\ell}{2\ell + 1}  j_{\ell - 1} (ks)
\right]
\nonumber\\
&=& 
(-i)^{\ell + 1} 
\sqrt{\pi(2\ell + 1)}
\int\frac{ds}{a(s)^2} e^{-\tau} \frac{j_\ell (ks)}{ds};
\nonumber \\ \nonumber \\
f_{\ell,m\ne0} (k\hat{z}) &=& 0.
\eea
Isotropy of the universe allows us to generalize above for $f_{\ell m}(\bold{k})$ with any $\hat{\bold{k}}$. 

\begin{figure*}
  \begin{center}
    \includegraphics[scale=0.75]{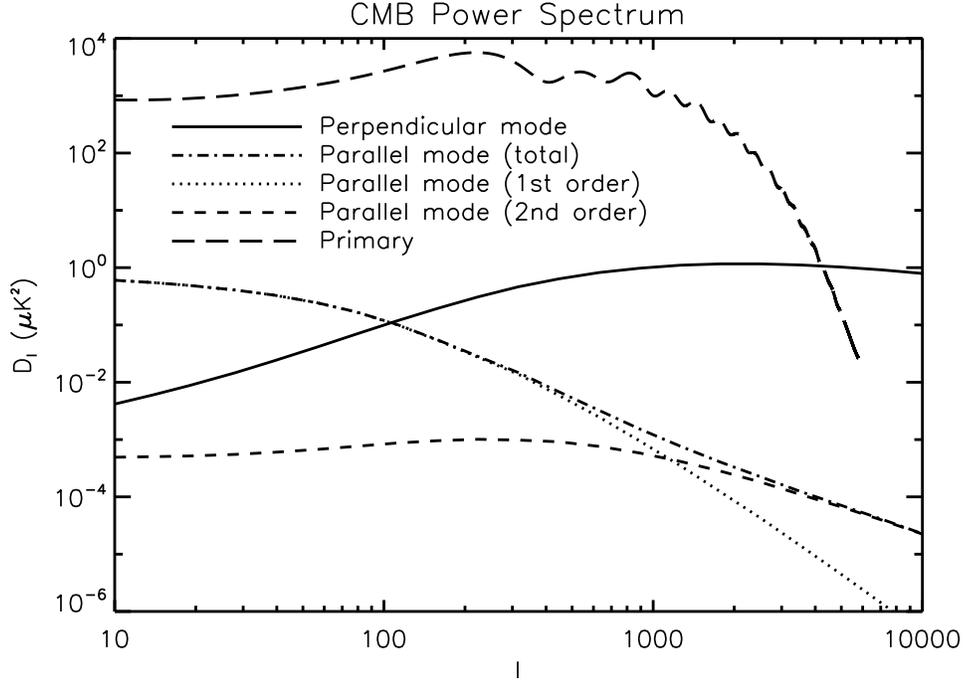}
  \caption{ Angular power spectrum of the kSZ effect from the transverse momentum field (solid) and the longitudinal momentum field (dot-dashed) in second order. The short-dashed and dotted lines denote contributions from the 1st and 2nd order terms in the longitudinal mode, respectively. The long-dashed line shows the primary CMB power spectrum. }
  \label{fig:D_l_par}
  \end{center}
\end{figure*}

Then, we obtain the final expression for the angular power spectrum:
\bea \label{eq:Clpar}
C^\parallel_\ell = \left(\frac{n_e \sigma_T}{c}\right)^2 \frac{1}{\pi} \int{\frac{ds}{a(s)^2}} \int{\frac{ds^\prime}{a(s^\prime)^2}} \int dk \frac{dj_\ell (ks)}{ds} \frac{dj_\ell (ks^\prime)}{ds^\prime} 
\sqrt{
P_{q_\parallel}(k,z(s))P_{q_\parallel}(k,z(s^\prime))
}.
\eea
%\left< \tilde{q}_\parallel(k,s) \tilde{q}^*_\parallel (k,s^\prime)\right>

We can show that $C^\parallel_\ell$ is much smaller than $C_\ell$ in Equation~(\ref{C_l}) using $P_{q_\perp}$ and $P_{q_\parallel}$ from the 2nd order calculation. For $P_{q_\perp}$, 2nd order is the leading order, and the expression (Eq.~\ref{eq:P_OV}) and the derivation are shown in Section~\ref{sec:basics}. For $P_{q_\parallel}$, the 2nd order terms follows the 1st order term:
\bea \label{eq:P_q_par}
P_{q_\parallel}(k,z) = \left( \frac{\dot{a}f}{k} \right)^2 P^{lin}_{\delta\delta}(k,z) + \dot{a}^2 f^2 \int \frac{d^3 k^\prime}{(2 \pi)^3} P^{lin}_{\delta\delta} (|\bold{k}-\bold{k^\prime}|,z) P^{lin}_{\delta\delta} (\bold{k^\prime},z) \frac{k\mu^\prime(k\mu^\prime-k^\prime {\mu^\prime}^2+k^\prime)}{{k^\prime}^2(k^2 + {k^\prime}^2-2kk^\prime \mu^\prime)}
\eea
\citep{ma/fry:2002}. Here, $P^{lin}_{\delta\delta}$ is the linear density power spectrum and $\mu^\prime = \hat{\bold{k}}\cdot \hat{\bold{k^\prime}}$. 

In Figure~\ref{fig:D_l_par}, we show the 2nd order calculation of kSZ angular power spectrum from longitudinal and transverse modes. 1st order term in the longitudinal modes dominates at $\ell \lesssim 100$, but diminishes rapidly in increasing $\ell$ as argued in \cite{vishniac:1987}. At $\ell \gtrsim 3000$ where the primary CMB vanishes to the level of allowing kSZ signal to be measured, the longitudinal mode contribution is below the transverse mode contribution by four orders of magnitude or more. Based on this comparison, we do not expect the longitudinal modes to be important even to a percent level and we shall ignore their contribution in this work.

\section{Connected Term in Perturbation Theory} \label{sec:P_q,c}

We take the third term in Equation~(\ref{eq:Pq}) and express it in terms
of $\theta \equiv \nabla \cdot \bold{v}$, which is equivalent to say $\tilde \bold{v}(\bold{k}) = i(\hat{\bold{k}}/k) \tilde{\theta}$:
\bea 
P_{q_\perp,c} &=& \int \frac{d^3k^\prime}{(2\pi)^3}\int\frac{d^3k^{\prime\prime}}{(2\pi)^3} 
[ \hat\bold{k^{\prime}}\cdot  \hat\bold{k^{\prime\prime}} - (\hat\bold{k} \cdot \hat\bold{k^{\prime}}) (\hat\bold{k} \cdot \hat\bold{k^{\prime\prime}}) ]
P_{\delta \delta v v,c}(\mathbf{k}-\mathbf{k^\prime},-\mathbf{k}-\mathbf{k^{\prime\prime}},\mathbf{k^\prime},\mathbf{k^{\prime\prime}}) 
\nonumber \\
&=&-(\dot{a}f)^2 \int \frac{d^3k^\prime}{(2\pi)^3}\int\frac{d^3k^{\prime\prime}}{(2\pi)^3}
\left[\frac{ \hat\bold{k^{\prime}}\cdot  \hat\bold{k^{\prime\prime}} - (\hat\bold{k} \cdot \hat\bold{k^{\prime}}) (\hat\bold{k} \cdot \hat\bold{k^{\prime\prime}}) }{k^\prime k^{\prime\prime}} \right]
P_{\delta\delta \theta \theta,c}(\mathbf{k}-\mathbf{k^\prime},-\mathbf{k}-\mathbf{k^{\prime\prime}},\mathbf{k^\prime},\mathbf{k^{\prime\prime}}).
\eea 
Let us derive $P_{\delta\delta \theta \theta,c}$ using perturbation theory.
We begin with the next-to-leading order expression for the full expression for $P_{\delta\delta \theta \theta}$ that includes the unconnected terms:
\bea
&&(2\pi)^3 P_{\delta\delta\theta\theta}(\mathbf{k_1},\mathbf{k_2},\mathbf{k_3},\mathbf{k_4}) \delta_{\rm D}(\mathbf{k_1}+\mathbf{k_2}+\mathbf{k_3}+\mathbf{k_4}) \nonumber\\
&=&\left<\tilde{\delta}(\mathbf{k_1}) \tilde{\delta}(\mathbf{k_2}) \tilde{\theta}(\mathbf{k_3}) \tilde{\theta}(\mathbf{k_4})\right> \nonumber\\
&=&\left<\tilde{\delta}^{(1)}(\mathbf{k_1}) \tilde{\delta}^{(2)}(\mathbf{k_2}) \tilde{\theta}^{(2)}(\mathbf{k_3}) \tilde{\theta}^{(1)}(\mathbf{k_4})\right> + \mbox{ cyclic (6 terms)} \nonumber \\
&+& \left<\tilde{\delta}^{(3)}(\mathbf{k_1}) \tilde{\delta}^{(1)}(\mathbf{k_2}) \tilde{\theta}^{(1)}(\mathbf{k_3}) \tilde{\theta}^{(1)}(\mathbf{k_4})\right>  +\mbox{ cyclic (4 terms)}.
\eea
The numbers in the superscripts indicate the perturbation order. We refer to the first case as $P_{1221}$ term and the second case as $P_{3111}$ term.
Note that above expression {\it is not} symmetric for switching one of $\bold{k_1}$ and $\bold{k_2}$ with one of $\bold{k_3}$ and $\bold{k_4}$ although it {\it is} symmetric for $\bold{k_1}\leftrightarrow \bold{k_2}$ and $\bold{k_3}\leftrightarrow \bold{k_4}$.

%$\left<\tilde{\delta}^{(1)}(\mathbf{k_1}) \tilde{\delta}^{(2)}(\mathbf{k_2}) \tilde{\theta}^{(2)}(\mathbf{k_3}) \tilde{\theta}^{(1)}(\mathbf{k_4})\right>$,~
%$\left<\tilde{\delta}^{(2)}(\mathbf{k_1}) \tilde{\delta}^{(1)}(\mathbf{k_2}) \tilde{\theta}^{(2)}(\mathbf{k_3}) \tilde{\theta}^{(1)}(\mathbf{k_4})\right>$,~

%$\left<\tilde{\delta}^{(2)}(\mathbf{k_1}) \tilde{\delta}^{(1)}(\mathbf{k_2}) \tilde{\theta}^{(1)}(\mathbf{k_3}) \tilde{\theta}^{(2)}(\mathbf{k_4})\right>,~
%\left<\tilde{\delta}^{(1)}(\mathbf{k_1}) \tilde{\delta}^{(2)}(\mathbf{k_2}) \tilde{\theta}^{(1)}(\mathbf{k_3}) \tilde{\theta}^{(2)}(\mathbf{k_4})\right>$.

One of $P_{1221}$ terms, $\left<\tilde{\delta}^{(1)}(\mathbf{k_1})
\tilde{\delta}^{(2)}(\mathbf{k_2}) \tilde{\theta}^{(2)}(\mathbf{k_3})
\tilde{\theta}^{(1)}(\mathbf{k_4})\right>$, is given by
\bea
&&\left<\tilde{\delta}^{(1)}(\mathbf{k_1}) \tilde{\delta}^{(2)}(\mathbf{k_2}) \tilde{\theta}^{(2)}(\mathbf{k_3}) \tilde{\theta}^{(1)}(\mathbf{k_4})\right>\nonumber\\
&=&
\int\frac{d^3\mathbf{k_{2,a}}}{(2\pi)^3} 
\int d^3\mathbf{k_{2,b}}
\int\frac{d^3\mathbf{k_{3,a}}}{(2\pi)^3}
\int d^3\mathbf{k_{3,b}}
\nonumber\\ 
&&
F^{(s)}_2(\mathbf{k_{2,a}},\mathbf{k_{2,b}})
G^{(s)}_2(\mathbf{k_{3,a}},\mathbf{k_{3,b}})
\delta_{\rm D}(\mathbf{k_2} - \mathbf{k_{2,a}} - \mathbf{k_{2,b}})
\delta_{\rm D}(\mathbf{k_3} - \mathbf{k_{3,a}} - \mathbf{k_{3,b}})
\nonumber\\ 
&&\left<\tilde{\delta}^{(1)}(\mathbf{k_1}) \tilde{\delta}^{(1)}(\mathbf{k_{2,a}}) \tilde{\delta}^{(1)}(\mathbf{k_{2,b}}) \tilde{\delta}^{(1)}(\mathbf{k_{3,a}})\tilde{\delta}^{(1)}(\mathbf{k_{3,b}}) \tilde{\delta}^{(1)}(\mathbf{k_4})\right>
\nonumber\\
&=&
\int\frac{d^3\mathbf{k_{2,a}}}{(2\pi)^3} 
\int d^3\mathbf{k_{2,b}}
\int\frac{d^3\mathbf{k_{3,a}}}{(2\pi)^3}
\int d^3\mathbf{k_{3,b}}
\nonumber\\ 
&&
F^{(s)}_2(\mathbf{k_{2,a}},\mathbf{k_{2,b}})
G^{(s)}_2(\mathbf{k_{3,a}},\mathbf{k_{3,b}})
\delta_{\rm D}(\mathbf{k_2} - \mathbf{k_{2,a}} - \mathbf{k_{2,b}})
\delta_{\rm D}(\mathbf{k_3} - \mathbf{k_{3,a}} - \mathbf{k_{3,b}})
\nonumber\\ 
&&
\left[
\left<\tilde{\delta}^{(1)}(\mathbf{k_1}) \tilde{\delta}^{(1)}(\mathbf{k_{2,a}})\right>
\left< \tilde{\delta}^{(1)}(\mathbf{k_{2,b}}) \tilde{\delta}^{(1)}(\mathbf{k_{3,a}})\right>
\left<\tilde{\delta}^{(1)}(\mathbf{k_{3,b}}) \tilde{\delta}^{(1)}(\mathbf{k_4})\right> \right.
\nonumber\\
&+&
\left<\tilde{\delta}^{(1)}(\mathbf{k_1}) \tilde{\delta}^{(1)}(\mathbf{k_{2,a}})\right>\left< \tilde{\delta}^{(1)}(\mathbf{k_{2,b}}) \tilde{\delta}^{(1)}(\mathbf{k_{3,b}})\right>\left<\tilde{\delta}^{(1)}(\mathbf{k_{3,a}}) \tilde{\delta}^{(1)}(\mathbf{k_4})\right> 
\nonumber\\
&+&
\left<\tilde{\delta}^{(1)}(\mathbf{k_1}) \tilde{\delta}^{(1)}(\mathbf{k_{2,a}})\right>
\left< \tilde{\delta}^{(1)}(\mathbf{k_{2,b}}) \tilde{\delta}^{(1)}(\mathbf{k_{4}})\right>
\left<\tilde{\delta}^{(1)}(\mathbf{k_{3,a}}) \tilde{\delta}^{(1)}(\mathbf{k_{3,b}})\right> ~~\mbox{(vanish)}
\nonumber\\
&+& \left<\tilde{\delta}^{(1)}(\mathbf{k_1})
\tilde{\delta}^{(1)}(\mathbf{k_{2,b}})\right> ~~\mbox{ (... are similar
to the above 3 lines)}
\nonumber\\
&+&
\left<\tilde{\delta}^{(1)}(\mathbf{k_1}) \tilde{\delta}^{(1)}(\mathbf{k_{3,a}})\right> 
\left<\tilde{\delta}^{(1)}(\mathbf{k_{2,a}}) \tilde{\delta}^{(1)}(\mathbf{k_{2,b}})\right> 
\left<\tilde{\delta}^{(1)}(\mathbf{k_{3,b}}) \tilde{\delta}^{(1)}(\mathbf{k_4})\right>  ~~\mbox{(vanish)}
\nonumber\\
&+&
\left<\tilde{\delta}^{(1)}(\mathbf{k_1}) \tilde{\delta}^{(1)}(\mathbf{k_{3,a}})\right> 
\left<\tilde{\delta}^{(1)}(\mathbf{k_{2,a}}) \tilde{\delta}^{(1)}(\mathbf{k_{3,b}})\right> 
\left<\tilde{\delta}^{(1)}(\mathbf{k_{2,b}}) \tilde{\delta}^{(1)}(\mathbf{k_4})\right> 
\nonumber\\
&+&
\left<\tilde{\delta}^{(1)}(\mathbf{k_1}) \tilde{\delta}^{(1)}(\mathbf{k_{3,a}})\right> 
\left<\tilde{\delta}^{(1)}(\mathbf{k_{2,a}}) \tilde{\delta}^{(1)}(\mathbf{k_4})\right> 
\left<\tilde{\delta}^{(1)}(\mathbf{k_{2,b}}) \tilde{\delta}^{(1)}(\mathbf{k_{3,b}})\right> 
\nonumber\\
&+& 
\left<\tilde{\delta}^{(1)}(\mathbf{k_1})
\tilde{\delta}^{(1)}(\mathbf{k_{3,b}})\right> ~~ \mbox{ (... are similar
to the above 3 lines)}
\nonumber\\
&+& 
\left.
\left<\tilde{\delta}^{(1)}(\mathbf{k_1})
\tilde{\delta}^{(1)}(\mathbf{k_4})\right> ~~(\mbox{... are the unconnected terms})
\right],
\eea
where 
$F_2^{(s)}(\bold{q_1},\bold{q_2})= \frac{5}{7}+\frac{1}{2}\frac{\bold{q_1}\cdot \bold{q_2}}{q_1 q_2}+\frac{2}{7} \left(\frac{\bold{q_1}\cdot \bold{q_2}}{q_1 q_2}\right)^2$ 
and 
$G_2^{(s)}(\bold{q_1},\bold{q_2})= \frac{3}{7}+\frac{1}{2}\frac{\bold{q_1}\cdot \bold{q_2}}{q_1 q_2}+\frac{4}{7} \left(\frac{\bold{q_1}\cdot \bold{q_2}}{q_1 q_2}\right)^2$.
The case that $\tilde \delta^{(1)}(\bold{k_1})$ is paired with $\tilde
\delta^{(1)}(\bold{k_{2,a}})$ is equivalent to the case that it is
paired with $\tilde \delta^{(1)}(\bold{k_{2,b}})$, the case that $\tilde
\delta^{(1)}(\bold{k_{2,a}})$ or $\tilde \theta^{(1)}(\bold{k_{3,a}})$
is paired with $\tilde \delta^{(1)}(\bold{k_{2,b}})$ or $\tilde
\theta^{(1)}(\bold{k_{3,b}})$, respectively, vanishes, and the case that
$\tilde \delta^{(1)}(\bold{k_1})$ is paired with $\tilde \delta^{(1)}(\bold{k_4})$ belongs to the unconnected moment. We proceed with the connected terms only:
\bea
&&\left<\tilde{\delta}^{(1)}(\mathbf{k_1}) \tilde{\delta}^{(2)}(\mathbf{k_2}) \tilde{\theta}^{(2)}(\mathbf{k_3}) \tilde{\theta}^{(1)}(\mathbf{k_4})\right>_c
\nonumber\\
&=&  
4
\int\frac{d^3\mathbf{k_{2,a}}}{(2\pi)^3} 
\int d^3\mathbf{k_{2,b}} 
\int\frac{d^3\mathbf{k_{3,a}}}{(2\pi)^3}
\int d^3\mathbf{k_{3,b}}
\nonumber\\ 
&&
F^{(s)}_2(\mathbf{k_{2,a}},\mathbf{k_{2,b}})
G^{(s)}_2(\mathbf{k_{3,a}},\mathbf{k_{3,b}})
\delta_{\rm D}(\mathbf{k_2} - \mathbf{k_{2,a}} - \mathbf{k_{2,b}})
\delta_{\rm D}(\mathbf{k_3} - \mathbf{k_{3,a}} - \mathbf{k_{3,b}})
\nonumber\\ 
&&
\left[
\left<\tilde{\delta}(\mathbf{k_1}) \tilde{\delta}(\mathbf{k_{2,a}})\right>
\left< \tilde{\delta}(\mathbf{k_{2,b}}) \tilde{\delta}(\mathbf{k_{3,a}})\right>
\left<\tilde{\delta}(\mathbf{k_{3,b}})
\tilde{\delta}(\mathbf{k_4})\right> 
\right.
\nonumber\\
&+&
\left.
\left<\tilde{\delta}(\mathbf{k_1}) \tilde{\delta}(\mathbf{k_{3,a}})\right> 
\left<\tilde{\delta}(\mathbf{k_{2,a}}) \tilde{\delta}(\mathbf{k_{3,b}})\right> 
\left<\tilde{\delta}(\mathbf{k_{2,b}}) \tilde{\delta}(\mathbf{k_4})\right> 
\right]
\nonumber\\ \nonumber \\
&=& 
4
\int\frac{d^3\mathbf{k_{2,a}}}{(2\pi)^3} 
\int        d^3\mathbf{k_{2,b}}
\int\frac{d^3\mathbf{k_{3,a}}}{(2\pi)^3}
\int        d^3\mathbf{k_{3,b}}
\nonumber\\ 
&&
F^{(s)}_2(\mathbf{k_{2,a}},\mathbf{k_{2,b}})
G^{(s)}_2(\mathbf{k_{3,a}},\mathbf{k_{3,b}})
\delta_{\rm D}(\mathbf{k_2} - \mathbf{k_{2,a}} - \mathbf{k_{2,b}})
\delta_{\rm D}(\mathbf{k_3} - \mathbf{k_{3,a}} - \mathbf{k_{3,b}})
\nonumber\\ 
&&
[
(2\pi)^3P(\mathbf{k_1})\delta_{\rm D}(\mathbf{k_1}      +\mathbf{k_{2,a}})
(2\pi)^3P(\mathbf{k_{2,b}})\delta_{\rm D}(\mathbf{k_{2,b}}+\mathbf{k_{3,a}})
(2\pi)^3P(\mathbf{k_4})\delta_{\rm D}(\mathbf{k_{3,b}}+\mathbf{k_4})
\nonumber\\
&+&
(2\pi)^3P(\mathbf{k_1})\delta_{\rm D}(\mathbf{k_1}      +\mathbf{k_{3,a}})
(2\pi)^3P(\mathbf{k_{2,a}})\delta_{\rm D}(\mathbf{k_{3,b}}+\mathbf{k_{2,a}})
(2\pi)^3P(\mathbf{k_4})\delta_{\rm D}(\mathbf{k_{2,b}}+\mathbf{k_4})
]
\nonumber\\ \nonumber \\
&=& 
4(2\pi)^3\delta_{\rm D}(\mathbf{k_1}+\mathbf{k_2}+\mathbf{k_3}+\mathbf{k_4})
\nonumber \\ 
&&
[
F^{(s)}_2(\mathbf{k_1}+\mathbf{k_2},-\mathbf{k_1})
G^{(s)}_2(\mathbf{k_1}+\mathbf{k_2},\mathbf{k_4})
P(\mathbf{k_1})
P(\mathbf{k_1}+\mathbf{k_2})
P(\mathbf{k_4})
\nonumber\\
&+&
G^{(s)}_2(\mathbf{k_1}+\mathbf{k_3},-\mathbf{k_1})
F^{(s)}_2(\mathbf{k_1}+\mathbf{k_3},\mathbf{k_4})
P(\mathbf{k_1})
P(\mathbf{k_1}+\mathbf{k_3})
P(\mathbf{k_4})
].
\eea

Similarly,
\bea
&&\left<\tilde{\delta}^{(2)}(\mathbf{k_1}) \tilde{\delta}^{(1)}(\mathbf{k_2}) \tilde{\theta}^{(2)}(\mathbf{k_3}) \tilde{\theta}^{(1)}(\mathbf{k_4})\right>_c \nonumber \\
&=&4(2\pi)^3\delta_{\rm D}(\mathbf{k_1}+\mathbf{k_2}+\mathbf{k_3}+\mathbf{k_4})
\nonumber\\
&&
[F^{(s)}_2(\mathbf{k_2}+\mathbf{k_1},-\mathbf{k_2})G^{(s)}_2(\mathbf{k_2}+\mathbf{k_1},\mathbf{k_4})
P(\mathbf{k_2})P(\mathbf{k_2} + \mathbf{k_1})P(\mathbf{k_4})
\nonumber\\ 
&+& 
G^{(s)}_2(\mathbf{k_2}+\mathbf{k_3},-\mathbf{k_2})F^{(s)}_2(\mathbf{k_2}+\mathbf{k_3},\mathbf{k_4})
P(\mathbf{k_2})P(\mathbf{k_2} + \mathbf{k_3})P(\mathbf{k_4})],
\eea
\bea
&&\left<\tilde{\delta}^{(1)}(\mathbf{k_1}) \tilde{\delta}^{(2)}(\mathbf{k_2}) \tilde{\theta}^{(1)}(\mathbf{k_3}) \tilde{\theta}^{(2)}(\mathbf{k_4})\right>_c \nonumber \\
&=&4(2\pi)^3\delta_{\rm D}(\mathbf{k_1}+\mathbf{k_2}+\mathbf{k_3}+\mathbf{k_4})
\nonumber\\
&&
[F^{(s)}_2(\mathbf{k_1}+\mathbf{k_2},-\mathbf{k_1})G^{(s)}_2(\mathbf{k_1}+\mathbf{k_2},\mathbf{k_3})
P(\mathbf{k_1})P(\mathbf{k_1} + \mathbf{k_2})P(\mathbf{k_3})
\nonumber\\ 
&+& 
G^{(s)}_2(\mathbf{k_1}+\mathbf{k_4},-\mathbf{k_1})F^{(s)}_2(\mathbf{k_1}+\mathbf{k_4},\mathbf{k_3})
P(\mathbf{k_1})P(\mathbf{k_1} + \mathbf{k_4})P(\mathbf{k_3})],
\eea
and
\bea
&&\left<\tilde{\delta}^{(2)}(\mathbf{k_1}) \tilde{\delta}^{(1)}(\mathbf{k_2}) \tilde{\theta}^{(1)}(\mathbf{k_3}) \tilde{\theta}^{(2)}(\mathbf{k_4})\right>_c \nonumber \\
&=&4(2\pi)^3\delta_{\rm D}(\mathbf{k_1}+\mathbf{k_2}+\mathbf{k_3}+\mathbf{k_4})
\nonumber\\
&&
[F^{(s)}_2(\mathbf{k_2}+\mathbf{k_1},-\mathbf{k_2})G^{(s)}_2(\mathbf{k_2}+\mathbf{k_1},\mathbf{k_3})
P(\mathbf{k_2})P(\mathbf{k_2} + \mathbf{k_1})P(\mathbf{k_3})
\nonumber\\ 
&+& 
G^{(s)}_2(\mathbf{k_2}+\mathbf{k_4},-\mathbf{k_2})F^{(s)}_2(\mathbf{k_2}+\mathbf{k_4},\mathbf{k_3})
P(\mathbf{k_2})P(\mathbf{k_2} + \mathbf{k_4})P(\mathbf{k_3})].
\eea

The following two terms can be expanded in the same way as well, but have slightly different forms:
\bea
&&\left<\tilde{\delta}^{(2)}(\mathbf{k_1}) \tilde{\delta}^{(2)}(\mathbf{k_2}) \tilde{\theta}^{(1)}(\mathbf{k_3}) \tilde{\theta}^{(1)}(\mathbf{k_4})\right>_c
\nonumber\\
&=& 
4(2\pi)^3\delta_{\rm D}(\mathbf{k_1}+\mathbf{k_2}+\mathbf{k_3}+\mathbf{k_4})
\nonumber\\
&&
[F^{(s)}_2(\mathbf{k_3}+\mathbf{k_1},-\mathbf{k_3})F^{(s)}_2(\mathbf{k_3}+\mathbf{k_1},\mathbf{k_4})
P(\mathbf{k_3})P(\mathbf{k_3} + \mathbf{k_1})P(\mathbf{k_4})
\nonumber\\ 
&+& 
F^{(s)}_2(\mathbf{k_3}+\mathbf{k_2},-\mathbf{k_3})F^{(s)}_2(\mathbf{k_3}+\mathbf{k_2},\mathbf{k_4})
P(\mathbf{k_3})P(\mathbf{k_2} + \mathbf{k_3})P(\mathbf{k_4})],
\eea

\bea
&&\left<\tilde{\delta}^{(1)}(\mathbf{k_1}) \tilde{\delta}^{(1)}(\mathbf{k_2}) \tilde{\theta}^{(2)}(\mathbf{k_3}) \tilde{\theta}^{(2)}(\mathbf{k_4})\right>_c
\nonumber\\
&=&4(2\pi)^3\delta_{\rm D}(\mathbf{k_1}+\mathbf{k_2}+\mathbf{k_3}+\mathbf{k_4})
\nonumber\\
&&[G^{(s)}_2(\mathbf{k_1}+\mathbf{k_3},-\mathbf{k_1})G^{(s)}_2(\mathbf{k_1}+\mathbf{k_3},\mathbf{k_2})
P(\mathbf{k_1})P(\mathbf{k_1} + \mathbf{k_3})P(\mathbf{k_2})
\nonumber\\
&+&
G^{(s)}_2(\mathbf{k_1}+\mathbf{k_4},-\mathbf{k_1})G^{(s)}_2(\mathbf{k_1}+\mathbf{k_4},\mathbf{k_2})
P(\mathbf{k_1})P(\mathbf{k_1} + \mathbf{k_4})P(\mathbf{k_2})
].
\eea

For $P_{3111}$ terms, we have
%Next, we expand one of $P_{3111}$ terms, $\left<\tilde{\delta}^{(3)}(\mathbf{k_1}) \tilde{\delta}^{(1)}(\mathbf{k_2}) \tilde{\theta}^{(1)}(\mathbf{k_3}) \tilde{\theta}^{(1)}(\mathbf{k_4})\right>$, as in following:
\bea
&&\left<\tilde{\delta}^{(3)}(\mathbf{k_1}) \tilde{\delta}^{(1)}(\mathbf{k_2}) \tilde{\theta}^{(1)}(\mathbf{k_3}) \tilde{\theta}^{(1)}(\mathbf{k_4})\right> 
\nonumber\\
&=& 
\int \frac{dk_{1,a}^3}{(2\pi)^3} \int \frac{dk_{1,b}^3}{(2\pi)^3} \int dk_{1,c}^3 ~
F_3(\mathbf{k_{1,a}},\mathbf{k_{1,b}},\mathbf{k_{1,c}})
\delta_{D}(\mathbf{k_{1}} - \mathbf{k_{1,a}} - \mathbf{k_{1,b}} - \mathbf{k_{1,c}})
\nonumber\\
&&
\left<\tilde{\delta}^{(1)}(\mathbf{k_{1,a}}) \tilde{\delta}^{(1)}(\mathbf{k_{1,b}}) \tilde{\delta}^{(1)}(\mathbf{k_{1,c}})   \tilde{\delta}^{(1)}(\mathbf{k_2}) \tilde{\delta}^{(1)}(\mathbf{k_3}) \tilde{\delta}^{(1)}(\mathbf{k_4})\right>
\nonumber\\
&=& 
\int \frac{dk_{1,a}^3}{(2\pi)^3} \int \frac{dk_{1,b}^3}{(2\pi)^3} \int dk_{1,c}^3 ~
F_3(\mathbf{k_{1,a}},\mathbf{k_{1,b}},\mathbf{k_{1,c}})
\delta_{D}(\mathbf{k_{1}} - \mathbf{k_{1,a}} - \mathbf{k_{1,b}} - \mathbf{k_{1,c}})
\nonumber\\
&&
\left<\tilde{\delta}^{(1)}(\mathbf{k_{1,a}}) \tilde{\delta}^{(1)}(\mathbf{k_{1,b}}) \tilde{\delta}^{(1)}(\mathbf{k_{1,c}})   \tilde{\delta}^{(1)}(\mathbf{k_2}) \tilde{\delta}^{(1)}(\mathbf{k_3}) \tilde{\delta}^{(1)}(\mathbf{k_4})\right>
\nonumber\\
&=& 
6(2\pi)^3F^{(s)}_3(\mathbf{-k_{2}},\mathbf{-k_{3}},\mathbf{-k_{4}})
P(\mathbf{k_2}) P(\mathbf{k_3}) P(\mathbf{k_4}) \delta_{D}(\mathbf{k_{1}} + \mathbf{k_{2}} + \mathbf{k_{3}} + \mathbf{k_{4}}),
\eea
\bea
&&\left<\tilde{\delta}^{(1)}(\mathbf{k_1}) \tilde{\delta}^{(3)}(\mathbf{k_2}) \tilde{\theta}^{(1)}(\mathbf{k_3}) \tilde{\theta}^{(1)}(\mathbf{k_4})\right> 
\nonumber\\
&=& 
6(2\pi)^3F^{(s)}_3(\mathbf{-k_{1}},\mathbf{-k_{3}},\mathbf{-k_{4}})
P(\mathbf{k_1}) P(\mathbf{k_3}) P(\mathbf{k_4}) \delta_{D}(\mathbf{k_{1}} + \mathbf{k_{2}} + \mathbf{k_{3}} + \mathbf{k_{4}}), \qquad 
\eea
\bea
&&\left<\tilde{\delta}^{(1)}(\mathbf{k_1}) \tilde{\delta}^{(1)}(\mathbf{k_2}) \tilde{\theta}^{(3)}(\mathbf{k_3}) \tilde{\theta}^{(1)}(\mathbf{k_4})\right> 
\nonumber\\
&=& 
6(2\pi)^3 G^{(s)}_3(-\mathbf{k_{1}},\mathbf{-k_{2}},\mathbf{-k_{4}})
P(\mathbf{k_1}) P(\mathbf{k_2}) P(\mathbf{k_4}) 
\delta_{D}(\mathbf{k_{1}} + \mathbf{k_{2}} + \mathbf{k_{3}} + \mathbf{k_{4}}), \qquad 
\eea
and
\bea
&&\left<\tilde{\delta}^{(1)}(\mathbf{k_1}) \tilde{\delta}^{(1)}(\mathbf{k_2}) \tilde{\theta}^{(1)}(\mathbf{k_3}) \tilde{\theta}^{(3)}(\mathbf{k_4})\right> 
\nonumber\\
&=& 
6(2\pi)^3 G^{(s)}_3(-\mathbf{k_{1}},\mathbf{-k_{2}},\mathbf{-k_{3}})
P(\mathbf{k_1}) P(\mathbf{k_2}) P(\mathbf{k_3}) \delta_{D}(\mathbf{k_{1}} + \mathbf{k_{2}} + \mathbf{k_{3}} + \mathbf{k_{4}}), \qquad 
\eea
where the kernels, $F_3$ and $G_3$, are given by the following recursion relations.
\bea
F_n(\bold{q_1},...,\bold{q_n}) &=&
\sum^{n-1}_{m=1}
\frac{G_m(\bold{q_1},...,\bold{q_m})}{(2n+3)(n-1)}
\left[
(2n+1)\alpha(\bold{q_1}+...+\bold{q_m},\bold{q_{m+1}}+...+\bold{q_n})
F_{n-m}(\bold{q_{m+1}},...,\bold{q_n})
\right. 
\nonumber \\  &&
\left.
+2\beta(\bold{q_1}+...+\bold{q_m},\bold{q_{m+1}}+...+\bold{q_n})
G_{n-m}(\bold{q_{m+1}},...,\bold{q_n})
\right],
\eea
\bea
G_n(\bold{q_1},...,\bold{q_n}) &=&
\sum^{n-1}_{m=1}
\frac{G_m(\bold{q_1},...,\bold{q_m})}{(2n+3)(n-1)}
\left[
3\alpha(\bold{q_1}+...+\bold{q_m},\bold{q_{m+1}}+...+\bold{q_n})
F_{n-m}(\bold{q_{m+1}},...,\bold{q_n})~~~~~~~~~~
\right. 
\nonumber \\  && 
\left.
+2n\beta(\bold{q_1}+...+\bold{q_m},\bold{q_{m+1}}+...+\bold{q_n})
G_{n-m}(\bold{q_{m+1}},...,\bold{q_n}) 
\right],
\eea
and $F^{(s)}_3$ and $G^{(s)}_3$ denote the symmetrized kernels of $F_3$ and $G_3$.

Combining results above and substituting them into
Equation~(\ref{eq:P_q,c}), we obtain the expression for the
transverse momentum power spectrum from the connected terms:
\bea \label{eq:P_q,c}
P^{\rm NLO}_{q_\perp,c} &=& - \int \frac{d^3k^\prime}{(2\pi)^3}\int\frac{d^3k^{\prime\prime}}{(2\pi)^3} 
\left[ \frac{  \hat\bold{k^{\prime}}\cdot  \hat\bold{k^{\prime\prime}}- (\hat\bold{k} \cdot \hat\bold{k^{\prime}}) (\hat\bold{k} \cdot \hat\bold{k^{\prime\prime}}) }{k^\prime k^{\prime\prime}}\right]
\nonumber \\
&&\left\{6\left[
F^{(s)}_3(\bold{k} + \bold{k^{\prime\prime}},-\bold{k^\prime} ,-\bold{k^{\prime\prime}} )
P(\bold{k} + \bold{k^{\prime\prime}}) P(\bold{k^\prime} ) P(\bold{k^{\prime\prime}} ) \right.\right.
\nonumber \\
&+& F^{(s)}_3(\bold{k} - \bold{k^\prime},\bold{k^\prime} ,\bold{k^{\prime\prime}} )
P(\bold{k} - \bold{k^\prime}) P(\bold{k^\prime} ) P(\bold{k^{\prime\prime}} )
\nonumber \\
&+&
G^{(s)}_3(\bold{k} - \bold{k^\prime},-\bold{k} - \bold{k^{\prime\prime}},\bold{k^{\prime\prime}} )
P(\bold{k} - \bold{k^\prime}) P(\bold{k} + \bold{k^{\prime\prime}}) P(\bold{k^{\prime\prime}} ) 
\nonumber \\
&+&
G^{(s)}_3(\bold{k} - \bold{k^\prime},-\bold{k} - \bold{k^{\prime\prime}},\bold{k^\prime} )
P(\bold{k} - \bold{k^\prime}) P(\bold{k} + \bold{k^{\prime\prime}}) P(\bold{k^\prime} )]
\nonumber \\
\nonumber \\
&+&
4[
F^{(s)}_2(\bold{k} - \bold{k^\prime},\bold{k^\prime}+ \bold{k^{\prime\prime}})
G^{(s)}_2(\bold{k^\prime}+\bold{k^{\prime\prime}},-\bold{k^{\prime\prime}})
P(\bold{k} - \bold{k^\prime})
P(\bold{k^\prime}+\bold{k^{\prime\prime}})
P(\bold{k^{\prime\prime}} )
\nonumber \\
&+&
F^{(s)}_2(\bold{k} - \bold{k^\prime},\bold{k^\prime}+\bold{k^{\prime\prime}})
G^{(s)}_2(-\bold{k^\prime}- \bold{k^{\prime\prime}},\bold{k^\prime})
P(\bold{k} - \bold{k^\prime})
P(\bold{k^\prime}+\bold{k^{\prime\prime}})
P(\bold{k^\prime})
\nonumber \\
&+&
G^{(s)}_2(\mathbf{k},-\bold{k} + \bold{k^\prime})
F^{(s)}_2(\mathbf{k},\bold{k^{\prime\prime}} )
P(\bold{k} - \bold{k^\prime})
P(\mathbf{k})
P(\bold{k^{\prime\prime}} )
\nonumber \\
&+&
G^{(s)}_2(\mathbf{k},-\bold{k} + \bold{k^\prime})
G^{(s)}_2(\mathbf{k},-\bold{k} - \bold{k^{\prime\prime}})
P(\bold{k} - \bold{k^\prime})
P(\mathbf{k})
P(\bold{k} + \bold{k^{\prime\prime}})
\nonumber \\
&+&
G^{(s)}_2(\bold{k} - \bold{k^\prime}+\bold{k^{\prime\prime}} ,-\bold{k} + \bold{k^\prime})
F^{(s)}_2(\bold{k} - \bold{k^\prime}+\bold{k^{\prime\prime}} ,\bold{k^\prime})
P(\bold{k} - \bold{k^\prime})
P(\bold{k} - \bold{k^\prime} + \bold{k^{\prime\prime}} )
P(\bold{k^\prime})
\nonumber \\
&+&
G^{(s)}_2(\bold{k} - \bold{k^\prime}+\bold{k^{\prime\prime}} ,-\bold{k} + \bold{k^\prime})
G^{(s)}_2(\bold{k} - \bold{k^\prime}+\bold{k^{\prime\prime}} ,-\bold{k} - \bold{k^{\prime\prime}})
P(\bold{k} - \bold{k^\prime})
P(\bold{k} - \bold{k^\prime} + \bold{k^{\prime\prime}} )
P(\bold{k} + \bold{k^{\prime\prime}})
\nonumber \\
&+&
F^{(s)}_2(-\bold{k^\prime}- \bold{k^{\prime\prime}},\bold{k} + \bold{k^{\prime\prime}})
G^{(s)}_2(-\bold{k^\prime}- \bold{k^{\prime\prime}},\bold{k^{\prime\prime}} )
P(\bold{k}+\bold{k^{\prime\prime}})
P(\bold{k^\prime}+\bold{k^{\prime\prime}})
P(\bold{k^{\prime\prime}} )
\nonumber \\
&+&
F^{(s)}_2(-\bold{k^\prime}- \bold{k^{\prime\prime}},\bold{k} + \bold{k^{\prime\prime}})
G^{(s)}_2(-\bold{k^\prime}- \bold{k^{\prime\prime}},\bold{k^\prime})
P(\bold{k}+\bold{k^{\prime\prime}})
P(\bold{k^\prime}+ \bold{k^{\prime\prime}})
P(\bold{k^\prime})
\nonumber \\
&+&
G^{(s)}_2(-\bold{k} - \bold{k^{\prime\prime}}+\bold{k^\prime},\bold{k} + \bold{k^{\prime\prime}})
F^{(s)}_2(-\bold{k} - \bold{k^{\prime\prime}}+\bold{k^\prime},\bold{k^{\prime\prime}} )
P(\bold{k} + \bold{k^{\prime\prime}})
P(\bold{k} - \bold{k^\prime}+ \bold{k^{\prime\prime}} )
P(\bold{k^{\prime\prime}} )
\nonumber \\
&+&
G^{(s)}_2(-\bold{k},\bold{k} + \bold{k^{\prime\prime}})
F^{(s)}_2(-\bold{k},\bold{k^\prime})
P(\bold{k} + \bold{k^{\prime\prime}})
P(\bold{k})
P(\bold{k^\prime})
\nonumber \\
&+&
F^{(s)}_2(\bold{k},-\bold{k^\prime})
F^{(s)}_2(\bold{k},\bold{k^{\prime\prime}} )
P(\bold{k})
P(\bold{k^\prime})
P(\bold{k^{\prime\prime}} )
\nonumber \\
&+&
\left. \left.
F^{(s)}_2(\bold{k} -\bold{k^\prime} + \bold{k^{\prime\prime}},  \bold{k^\prime})
F^{(s)}_2(\bold{k} -\bold{k^\prime} + \bold{k^{\prime\prime}}, -\bold{k^{\prime\prime}} )
P(\bold{k^\prime})
P(\bold{k} -\bold{k^\prime} + \bold{k^{\prime\prime}})
P(\bold{k^{\prime\prime}} )
\right]\right\}
\eea

\section{Unconnected Term in Perturbation Theory}  \label{sec:P_q,uc}

For the unconnected term, we begin by substituting the next-to-leading order power spectra of $P^{(2)}_{\delta \delta}$, $P^{(2)}_{\delta \theta}$ and $P^{(2)}_{\theta \theta}$ in Equation~(\ref{eq:MF}). Then, perturbation theory gives
\bea 
P^{(2)}_{\delta \delta}(\bold{k}) &=& \int \frac{d^3k^\prime}{(2\pi)^3} F^{(s)}_2 (\bold{k}-\bold{k^\prime},\bold{k^\prime})^2 P(|\bold{k}-\bold{k^\prime}|) P(k^\prime)
\nonumber \\
&+& 
2\int \frac{d^3k^\prime}{(2\pi)^3} F^{(s)}_3 (\bold{k},-\bold{k^\prime},\bold{k^\prime}) P(k) P(k^\prime),
\nonumber \\ 
P^{(2)}_{\delta \theta}(\bold{k}) &=& \int \frac{d^3k^\prime}{(2\pi)^3} F^{(s)}_2 (\bold{k}-\bold{k^\prime},\bold{k^\prime}) G^{(s)}_2 (\bold{k}-\bold{k^\prime},\bold{k^\prime}) P(|\bold{k}-\bold{k^\prime}|) P(k^\prime)
\nonumber \\
&+& 
\int \frac{d^3k^\prime}{(2\pi)^3} F^{(s)}_3 (\bold{k},-\bold{k^\prime},\bold{k^\prime}) P(k) P(k^\prime)
\nonumber \\
&+& 
\int \frac{d^3k^\prime}{(2\pi)^3} G^{(s)}_3 (\bold{k},-\bold{k^\prime},\bold{k^\prime}) P(k) P(k^\prime),
\nonumber \\ 
 P^{(2)}_{\theta\theta}(\bold{k}) &=& 
 \int \frac{d^3k^\prime}{(2\pi)^3} G^{(s)}_2 (\bold{k}-\bold{k^\prime},\bold{k^\prime})^2 P(|\bold{k}-\bold{k^\prime}|) P(k^\prime)
\nonumber \\
&+&
2 \int \frac{d^3k^\prime}{(2\pi)^3} G^{(s)}_3 (\bold{k},-\bold{k^\prime},\bold{k^\prime}) P(k) P(k^\prime).
\eea
Substituting $P_{\delta \delta}=P^{(2)}_{\delta \delta}$, $P_{\delta
\theta}=P^{(2)}_{\delta \theta}$ and $P_{\theta \theta}=P^{(2)}_{\theta
\theta}$ to Equation~(\ref{eq:MF}) and dropping higher order terms like $P^{(2)}P^{(2)}$, 
we obtain the expression for the
transverse momentum power spectrum from the unconnected terms:
\bea \label{eq:P_q,uc} 
&&P^{\rm{NLO}}_{q_{\perp},uc}(k,z) = 
\dot{a}^2f^2\int \frac{d^3k^\prime}{(2\pi)^3} (1-{\mu^\prime}^2) 
\nonumber \\ 
&&
\left\{ \frac{1}{{k^\prime}^2}\left[
P(|\bold{k}-\bold{k^\prime}|) P^{(2)}_{\theta\theta} (k^\prime) +
P^{(2)}_{\delta\delta}(|\bold{k}-\bold{k^\prime}|) P (k^\prime)\right] +
\right.
\nonumber \\ 
&&\left.
\frac{1}{|\bold{k}-\bold{k}^\prime|^2}
\left[P(|\bold{k}-\bold{k^\prime}|)
P^{(2)}_{\delta \theta} (k^\prime)
+
P^{(2)}_{\delta \theta}(|\bold{k}-\bold{k^\prime}|)
P (k^\prime)\right]
\right\}.~~~~~
\eea

\bibliographystyle{apj}
\end{CJK}
\bibliography{reference}
\end{document}